\documentclass{aa}

\usepackage[utf8]{inputenc}

\usepackage{newtxtext,newtxmath}
\usepackage{graphicx}
\usepackage{color}
\usepackage{subfigure}
\usepackage{mathtools}
\usepackage{amsmath}
\usepackage{lastpage}

\def\me{$\,{\rm M}_{\oplus}\,$}

\usepackage{ulem,bm}
\definecolor{mygray}{gray}{0.6}
\definecolor{orange}{rgb}{1,0.4,0}

\begin{document} 

\title{How planets form by pebble accretion \\ 
{V. Silicate rainout delays contraction of sub-Neptunes} \\
}
\titlerunning{Evolution of rocky planets formed by pebble accretion}
\author{Allona Vazan\inst{1} \and Chris W. Ormel\inst{2} \and Marc G. Brouwers\inst{3} }
\institute{Astrophysics Research Center (ARCO), Department of Natural Sciences, The Open University of Israel, Raanana 4353701, Israel 
\and Department of Astronomy, Tsinghua University, 30 Shuangqing Rd, 100084 Beĳing, China
\and Institute of Astronomy, University of Cambridge, Madingley Road, Cambridge CB3 0HA, UK}

\abstract
{The characterization of Super-Earth-to-Neptune sized exoplanets relies heavily on our understanding of their formation and evolution. In this study, we link a model of planet formation by pebble accretion \citep{ormel21} {to the planets' long-term observational properties by calculating the} interior evolution, starting from the dissipation of the protoplanetary disk. We investigate the evolution of the interior structure in 5--20\me planets, {accounting for} silicate redistribution caused by convective mixing, rainout (condensation and settling), and mass loss. 
Specifically, we have followed the fate of the hot silicate vapor that remained in the planet's envelope after planet formation, as the planet cools.
We find that disk dissipation is followed by a rapid contraction of the envelope from the Hill/Bondi radius to about one-tenth of that size within 10 Myr. 
Subsequent cooling leads to substantial growth of the planetary core through silicate rainout, accompanied by inflated radii, in comparison to the standard models of planets that formed with core-envelope structure. 
We examine {the dependence of rainout on} the planet's envelope mass, distance from its host star, its silicate mass, and the atmospheric opacity.
We find that the population of planets formed with polluted envelopes can be roughly divided in three groups, based on the mass of their gas envelopes: bare rocky cores that have shed their envelopes, super-Earth planets with a core-envelope structure, and Neptune-like planets with diluted cores that undergo gradual rainout. For polluted planets formed with envelope masses below 0.4\me, we anticipate that the inflation of the planet's radius caused by rainout will enhance mass loss by a factor of 2--8 compared to planets with non-polluted envelopes.  
Our model provides an explanation for bridging the gap between the predicted composition gradients in massive planets and the core-envelope structure in smaller planets.}

\maketitle

\keywords{Planets and satellites: formation --
             Planets and satellites: interiors --
             Planets and satellites: composition --}

\section{Introduction}
The mass-radius relation of exoplanets (i.e., bulk density) depends on their composition, but also on composition distribution in the interior. Here, mixtures of elements are usually more compact than separate composition layers \citep[e.g.,][]{baraffe08,Fortney13,dornlich21}. 
In particular, the mass-radius relation of silicate-dominated planets with only a few percent of gas are more sensitive to the thermal evolution of the metal-rich interior \citep{lopezfor14,vazan18a,vazan18c} and to the mean molecular weight (pollution) of the gas envelope.  

{Observations from online and upcoming facilities as} JWST \citep{greene16jwst}, PLATO \citep{rauer14short} and ARIEL \citep{tinetti18short} space telescopes are expected to provide high quality data as well as improved statistics for stellar ages (PLATO) and atmospheric compositions (ARIEL, JWST). 
One way to shed light on the nature of these planets is by studying the sequence of planet formation and long term thermal evolution. While planet formation sets the initial mass and composition of the planet, the long term evolution connects it to the observed properties. 

Planet formation models that follow the solid deposition in the interior show that most of the accreted metals are distributed in the gas envelope once the planet exceeds about 2\me, due to solid-gas interaction in the growing envelope \citep{boden18,brouwers18,valletta19,brouwers20,ormel21,steinmeyer23}.
Thus, formation-evolution of 5-20\me rocky planets involve the evolution of polluted envelopes, usually with non-uniform distribution of metals in it. 

A question {that arises is} whether such a {non-uniform} structure is stable {on timescales where these planets are observed -- i.e., several Gyr --} {or whether the internal state has been re-arranged into a standard core-envelope structure ("standard" refers to how these planets are typically perceived) or something in between (i.e., compositional layers)}. 
Processes in the interior like mixing or settling redistribute {matter} along with the thermal evolution of the planet. If convection is vigorous {\it convective-mixing} can erode the initial metal distribution and homogenize it \citep{vazan15}. Alternatively, progressive cooling can lead to over-saturation in the polluted envelope and {\it rainout} (condensation and settling) of {silicate} droplets to deeper hotter layers \citep{brouwers20}. Thus, the planetary structure {directly} after formation is not necessarily the structure {at the present-day}\footnote{Compositional diffusion might also change the planetary structure, but is found to be negligible in comparison to convective-mixing and rainout, as is shown in section~\ref{sec:times}.}.

Evolution models usually use arbitrary initial conditions. While this is a fair assumption for interior models of 2-3 homogeneous composition layers, it cannot hold for evolution of planets with non-uniform composition distribution. When composition in planets after their formation is not uniformly distributed, evolution is not adiabatic \citep{ledoux47,rosenblum11}, and hence thermal evolution is sensitive to initial conditions (i.e., formation outcomes). 

In \citet{ormel21} we found that rocky 5-20\me planet formed by pebble accretion have a typical structure of a small rocky core surrounded by a silicate vapor composition gradient and a vapor-rich (>40\%) convective envelope on top of it. In {that} work {\it Phase IV} {was defined as} the phase where the protoplanetary disk dissipates, the outer layers of the atmosphere may evaporate, and where progressive cooling may cause the vapor to rain out. Here we aim to model this phase.
 
In a recent Letter \citep{vazanormel23} we show that condensation and settling (rainout) of silicates in the envelope of sub-Neptune planets have a noticeable effect on their observed radii. In this paper we generalize this study to a wider range of planetary masses and conditions, and denote the trends in evolution of polluted envelopes.  
We calculate the long term properties of 5-20\me rocky planets composed of silicate, hydrogen and helium, that formed via pebble accretion. The model contains: (1) initial interior structure from planet formation model, (2) thermal evolution including relevant heat transport mechanisms and energy sources, (3) structural evolution (material transports) by convective-mixing and settling, (4) gas mass loss from young atmospheres. 

\section{Method}\label{sec:method}

\subsection{Formation-Evolution interface}

Our initial models are the resulting interior models of \cite{ormel21} for planet formation by rocky pebble accretion. At the end of the formation phase the young planet is embedded in the disk, and its radius is of the order of hundred Earth radii (Bondi radius / Hill radius). 
The interior of the young planet is typically structured in four stable layers: a small rocky core (1-2\me), a silicate vapor composition gradient, a uniform silicate-rich convective envelope, and a thin silicate-poor saturated atmosphere.

\begin{figure*}
\sidecaption
\includegraphics[width=0.7\textwidth]{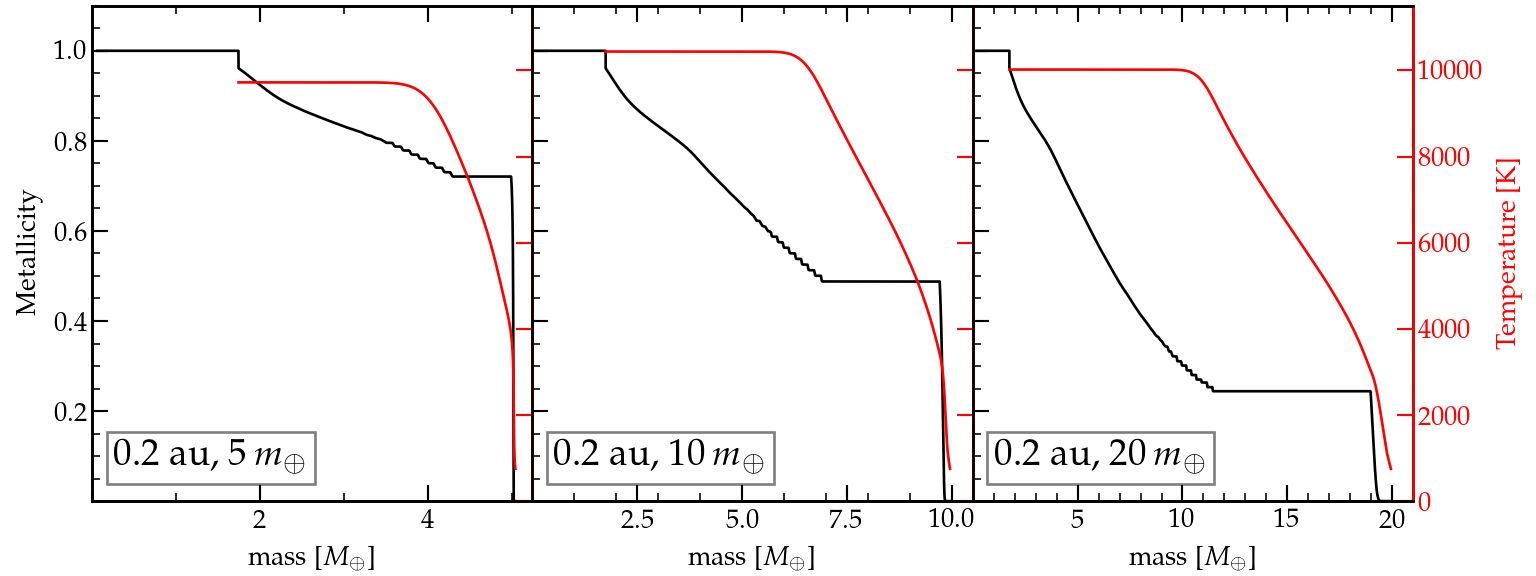}
\caption{Post-formation outcome of interior structure of planets assembled by pebble accretion \citep{ormel21}. Shown are the silicate mass fraction distribution (black) and the envelope temperature profile (red) in planets of 5\me (left), 10\me (middle), and 20\me (right), formed at 0.2\,AU form a Sun-like star.}\label{fig:form}
\end{figure*}

In Figure~\ref{fig:form} we present the silicate distribution (Z) and temperature profile of planets at the end of their formation phase. The formation models in the figure are calculated for accretion of 1\,mm rocky ($SiO_2$) pebbles, at a solid accretion rate of $10^{-5}$ \me\ yr$^{-1}$, where opacity in the formation phase is of gas and pebbles. 
In section~\ref{sec:pars} we discuss how varying these parameters affects the results. The parameters from the formation model that we use in the evolution calculation are interior profile of temperature, density, luminosity, entropy, and composition (species mass fraction).  

{In our simulation the transition from disk phase to post-disk long term evolution is sharp, as the formation outcome model is the input of the evolution model. {Specifically, the pressure at the outer boundary is now given by the photospheric conditions, which is significantly lower than the disk pressure.} The temperature at the outer boundary is determined by the distance from the (sun-like) star, which is unchanged between formation and evolution.}

For consistency of the thermal evolution model with the planet formation model we use the same set of equations of state for hydrogen and helium \citep{scvh}, and $SiO_2$ rock \citep{faik18}. 
Irradiation by the parent star is for the same location as in the disk phase (i.e., no migration). We evolve formation models of planets at distances of 0.2\,AU and 1\,AU from a sun-like star\footnote{Planets at larger distances are expected to contain significant amount of volatiles and thus are beyond the scope of this work.}. The number of mass grid points is reduced by interpolation to 500, which is optimal for the thermal evolution code runs.

\subsection{Thermal evolution model}

The interior thermal evolution code is a Henyey-like\footnote{An iterative implicit integration method for boundary value problems, in which the structure equations are solved together with the evolution equations \citep{henyey64}.} code that solves the set of structure-evolution equations simultaneously on an adaptive grid of mass from center to surface \citep{kovetz09}. Thus, the thermal state of the deep silicate core, including its compression and cooling is integral part of the evolution scheme. 
{At} each mass grid-point we obtain a solution for the equations of continuity (\ref{eq:cont}), hydrostatic equilibrium (\ref{eq:hydro}), energy transport via convection, radiation, and conduction (3), energy balance (\ref{eq:ebalance}), and composition flux (5): 
\begin{equation}\label{eq:cont}
    \frac{\partial}{\partial m}\frac{4\pi}{3}r^3=\frac{1}{\rho}
\end{equation}

\begin{equation}\label{eq:hydro}
    \frac{\partial p}{\partial m}=-\frac{Gm}{4\pi r^4}
\end{equation}

\begin{subequations}
\begin{equation}\label{eq:etrans}
    \frac{\partial \ln T}{\partial m}=\nabla \frac{\partial \ln p}{\partial m}
\end{equation}
where the symbols $r,m,\rho,p,T$ are the radius, mass, density, pressure and temperature respectively. 
$\nabla = d\ln T / d\ln P$ is the temperature gradient determined by the heat transport mechanism according to the Ledoux convection criterion \citep{ledoux47}: 

\begin{equation}
        \nabla = 
\begin{cases}
    \nabla_\mathrm{R} & \text{  for  } \nabla_\mathrm{R} \leq \nabla_{\rm ad}+\nabla_{\rm X}\\
    \nabla_{\rm MLR}  & \text{  for  } \nabla_{\rm R} > \nabla_{\rm ad}+\nabla_{\rm X}
\end{cases}
\end{equation}
$\nabla_{\rm ad}$ is the adiabatic gradient, 
\begin{equation}
    \nabla_{\rm ad}=\frac{d\ln T}{d\ln p}\Bigg{|}_s
\end{equation}
$\nabla_X$ is the composition dependent gradient, 

\begin{equation}
    \nabla_X=\sum_j \frac{\partial \ln T}{\partial X_j}\frac{dX_j}{d \ln p}
\end{equation}
where $X_j$ is mass fraction of the $j$th species. 

$\nabla_R$ is the radiative (and/or conductive) temperature gradient, 
\begin{equation}
    \nabla_{\rm R}=\frac{\kappa_{\rm op} L}{4\pi cGm}\frac{p}{4p_R}
\end{equation}
$p_{\rm R}$ is the radiation pressure, $L$ luminosity, $\kappa_{\rm op}$ is the harmonic mean of radiative ($\kappa_{\rm rad}$) and conductive ($\kappa_{\rm cond}$) opacities: 

    \begin{equation}\label{eq:opac}
       \kappa_{\rm op}=\left( \frac{1}{\kappa_{\rm rad}} + \frac{1}{\kappa_{\rm cond}}\right)^{-1}
    \end{equation}
\end{subequations}

Grains or pebbles are not expected to remain in the upper envelope after the formation phase \citep{brouwers21,movsh10,ormel14,mordasini14}, and therefore we use the radiative opacity of \cite{freedman14} for grain-free atmospheres.   
{The opacity in each layer is calculated for the layer’s given pressure-temperature-metallicity values. However, since in practice atmospheric metallicity composed of various elements, and not solely of silicate vapor that we consider in this work, we apply the condition 
\begin{equation}\label{eq:Zeff}
    Z_\textrm{radiative} = \max(Z,Z_\textrm{min})
\end{equation} 
{as} the metallicity for which the Freedman opacities are calculated. {Here, we take $Z_\mathrm{min}$ equal to the solar metallicity by default. Note that while $Z_\mathrm{min}$ is fixed throughout the simulation, $Z$ is calculated based on the L-V curve (eq.~\ref{eq:sat}). Thus, in layers where the silicate vapor fraction drops below solar the solar metallicity is adopted for the calculation of the opacity.}}
Conduction is governed by the {silicate} component, as molecular hydrogen is poor conductor \citep{french12} and pressure-temperature in interiors of super-Earths are below the metallic hydrogen level. 
Conductive opacity is obtained by extrapolation of Earth rock parameters to high pressure and temperature as described in {appendix~\ref{sec:app2}.}

$\nabla_{\rm MLR}$ is the temperature gradient calculated by the Mixing Length Recipe (MLR) \citep{mihalas78,kippenhahn13}, where turbulent Rayleigh-Bénard convection is simulated in 1D. We take the size of a convective cell to be half the local scale height\footnote{See \cite{vazan15} for more details on the MLR model and on the effect of its parameters on results.}. 

The energy balance in the interior is according to 
\begin{equation}\label{eq:ebalance}
    \frac{\partial u}{\partial t}+p\frac{\partial}{\partial t} \frac{1}{\rho}=q-\frac{\partial L}{\partial m}
\end{equation}
where specific energy ($u$), contraction, and contribution of radiogenic heating in the rock ($q$), are considered. 

\begin{subequations}
Material transport by convective mixing is calculated as a diffusive-convective flux of particles \citep{kippenhahn13}:
\begin{equation}\label{eq:compF}
     \frac{\partial Y_j}{\partial t}= \frac{\partial}{\partial m} F_j
\end{equation}
where $Y_j=X_j/A_j$ is the number fraction of the $j$th element, namely the mass fractions divided by the atomic mass of the element, and $F_j$ is the particle flux of the $j$th element: 

\begin{equation}
    F_j=\sigma_j \frac{\partial Y_j}{\partial m}
\end{equation}

\begin{equation}
    \sigma_j=\left(\frac{dm}{dr}\right)^2 v_c l_c
\end{equation}
The particle flux is determined by a diffusive coefficient $\sigma_j$, which depends on the convection velocity $v_c$ and mixing length $l_c$. 
\end{subequations}

At every mass layer and timestep the equation of state for a mixture of the three materials - hydrogen, helium (H, He) and {silicate} ($SiO_2$) - is calculated according to the additive volume law
    \begin{equation}
        \rho = \left( \sum_j \frac{X_j}{\rho_j}\right)^{-1}
    \end{equation}
where $\rho_j$ is the density of the $j$th species, obtained from the {equation of state} (H,He from \cite{scvh} and $SiO_2$ from \cite{faik18}). The other thermodynamic properties of the mixture are calculated as described in \cite{vazan13}.

\subsubsection{Rainout model}

As the planet cools the vapor in the envelope become oversaturated, and part of it condensates and settles down to deeper hotter (undersaturated) layers. 
The processes sweeps the {silicate} from the outer layers downwards and gives rise to late core growth. The rainout phase terminates when all silicates {have} settled into a silicate core surrounded by a metal-poor hydrogen-helium envelope\footnote{We ignore residuals of less than 2\% in mass that may stay in the envelope.}. Once the core-envelope structure has {been established}, the structure is static and further cooling doesn't affect the composition distribution. 
We define the rainout timescale {$t_\mathrm{rain}$} as the time from disk dissipation until all silicates have settled and a core-envelope structure has been {established}.

Settling from an supersaturated layer into a deeper undersaturated layer is assumed to be instantaneous, since the settling timescale of a {silicate} droplet is orders of magnitude smaller than the evolution timescale (see Section~\ref{sec:times} below). {For the same reason} we exclude the possibility of supersaturation.
We model rainout numerically by removing the excess of {silicate} (the amount above the saturation metallicity) from a layer $i$ to the layer below it $i-1$. 
Between timestep $t$ and $t+1$, whenever the pressure-temperature in the layer ($p_i, T_i$) are such that $Z_i>Z_{sat(p_i,T_i)}$ we {take}:
\begin{subequations}
    \begin{equation}
        Z_i^{t+1}=Z_{sat(p_i,T_i)}
    \end{equation}
and the excess {amount} of {silicate} $\Delta m_Z = (Z_i-Z_{sat(p_i,T_i)}) m_i$ is moved from the layer $i$ to the layer below it, enhancing its metallicity:
    \begin{equation}
        Z_{i-1}^{t+1}=\frac{Z_{i-1}^{t} m_{i-1} +\Delta m_Z}{m_{i-1}}
    \end{equation}
\end{subequations}

\begin{figure}
\centerline{\includegraphics[width=\columnwidth]{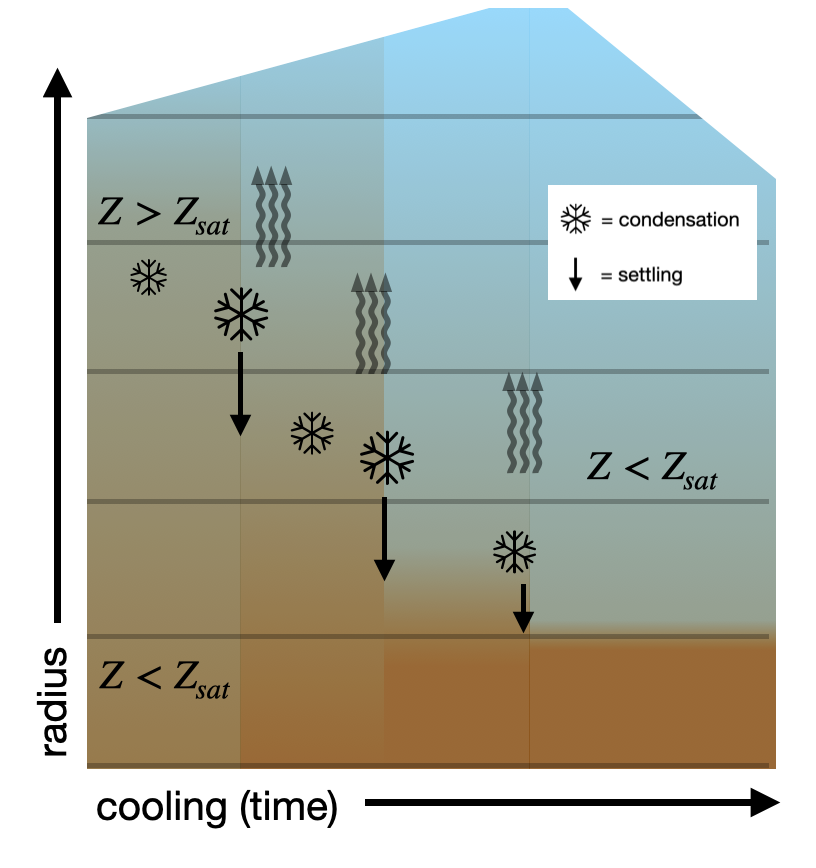}}
\caption{Schematic picture of the rainout process in a polluted envelope. Cooling (from left to right) leads to condensation of silicate in oversaturated outer envelope and settling of the condensed silicates to deeper undersaturated layers. Condensation and settling release energy (curly arrows) that leads to envelope extension and inflated radius in comparison to a planet with same composition in core-envelope structure. Rainout {ends} once the planet reaches a core-envelope structure, and the radius inflation is diminished. Axes in the figure are not to scale.}\label{fig:rain}
\end{figure}

The rainout of {silicate} releases energy -- latent heat of condensation and gravitational energy of settling. 
{An additional energy source is the late release of the {heat from} formation that is {liberated after the compositional gradient has been erased}}. This energy is more difficult to estimate as it varies with formation conditions. 
The energy change by rainout is included in our equation scheme: the energy of condensation (latent heat) is part of the equation of state, the energy of settling (gravitational energy) is obtained by the structure equations, and the release of heat from formation is calculated self-consistently by the evolution model. 
A schematic sketch of the rainout process and its energetic contribution to the thermal evolution appear in figure~\ref{fig:rain}. 

The saturation criterion is defined by silicate liquid-vapor (L-V) curve \citep[e.g.][]{podolak88,boden18,stevenson22}. For consistency with our formation model \citep{ormel21} we use the fit of L-V curve (lower branch) of \cite{kraus12}: 
\begin{equation}\label{eq:sat}
    \rho_{\rm sat}=\exp \left( a-\frac{b}{T}+\left(\frac{T}{T_{\rm c}}\right)^c \right)
\end{equation}
in cgs units, where $a = 9.0945$, $b = 5.630\times 10^4$\,K, $c = 13.26$, and $T_{\rm c} = 5130$\,K. 

\subsubsection{Atmosphere model and mass loss estimate}\label{sec:atmos}

The atmosphere in our model is gray and plane-parallel, and the temperature distribution is calculated for a vertical (maximal) irradiation flux, with no angle dependency of the incident flux \citep[e.g.,][]{guillot10}. {The temperature in the atmosphere} follows:
\begin{equation}\label{eq_bc6}
    \sigma_{SB} T^4(\tau)=\left(\frac{3}{4}\tau+\frac{1}{4}\right)F_E+g(\tau)\sigma_{SB} T_{\rm irr}^4.
\end{equation}
where $\tau$ is optical depth, $F_E$ {the outgoing} energy flux, $g(\tau)=\frac{3}{2}(1-\frac{1}{2} e^{-\tau})$ \citep{kanikovetz67}, and $\sigma_{SB}$ is Stefan-Boltzmann's constant.

The atmospheric boundary is the planetary photosphere, with optical depth $\tau_s=1$. The net outward luminosity $L_s$ from the surface of the planet $R_p$ is
\begin{equation}\label{eq_bc7}
    L_s = 4\pi R_p^2F_E = 4\pi R_p^2\sigma_{SB} \left[T^4-g(\tau_S) T_{\rm irr}^4\right].
\end{equation}

The irradiation temperature as a function of the distance from the star is 
\begin{equation}\label{eq:tirrd}
    T_{\rm irr}=\left(\frac{L_{\star}(1-A)}{16\pi\sigma d^2}\right)^{1/4}
\end{equation}
where $L_{\star}$ is the {stellar} luminosity, here taken to be solar, $d$ is the planet-star orbital separation, and $A$ is the bond albedo.

Mass loss is simulated as material escape from the outermost layer in a given rate. The outermost layer is poor in {silicate} vapor, due to efficient rainout and atmospheric temperatures well below {silicate} evaporation ($T_\mathrm{evap} \approx 2500\,K$). Therefore the mass loss (removal) is assumed to be of hydrogen-helium. 

\subsection{Analytical estimate for the rainout timescale}\label{sec:anal}

\cite{brouwers20} analytically estimated rainout timescales for super-Earth planets. Their calculation {assumes an} ideal gas equation of state, constant opacity and luminosity, and adiabatic cooling. It is calculated for a three layers structure of a constant density core, an intermediate uniform polluted envelope, and an outer hydrogen envelope. 

Here, we adopt a slightly modified form of the estimate of \cite{brouwers20} for the sedimentation timescale. We estimate the energy available for rainout as 
\begin{equation}
    \label{eq:new-Urain}
    U_\mathrm{rain} = \left( \frac{3G}{5} \max \left[ \frac{M_c}{r_c}, \frac{M_z}{r_z} \right] + u_\mathrm{lat}\right)(M_z-M_c)
\end{equation}
where $M_c$ and $M_z$ are the core mass and total heavily element mass, respectively, and $r_c$ and $r_z$ are the radii corresponding to these masses and $u_\mathrm{lat}$ is the latent heat per unit mass of silicate (SiO$_2$) condensation. In \citet{brouwers18} and \citet{ormel21} it has been shown that $M_c=2\,M_\oplus$ is an appropriate choice for the (solid) core mass after formation; the heavy element mass above this point ($M_z-M_c$) will be vaporized. In Eq.~(\ref{eq:new-Urain}) the two cases correspond to the limits where $M_z$ is small, in which case the gravitational potential is determined by the $M_c$, and where $M_z$ is large, respectively. The sedimentation luminosity follows \cite{brouwers20}:
\begin{equation}
    L_\mathrm{rain} = \frac{64\pi\sigma T_\mathrm{rcb}^3 T_\mathrm{vap}}{\kappa \rho_c} r_c^\frac{\gamma-2}{\gamma-1} r_B^\frac{1}{\gamma-1} \frac{M_c}{M_\mathrm{xy}} 
\end{equation}
where $T_\mathrm{rcb}$ is the radiative-convective boundary (RCB) temperature, set to be the evaporation temperature $T_\mathrm{vap}$, $\kappa$ is the radiative opacity at the radiative atmosphere, $\rho_c$ is the core density, and $M_\mathrm{xy}$ is the gas (H,He) mass. 
$r_B'$ is the modified Bondi radius
\begin{equation}
    r_B' =\frac{\gamma-1}{\gamma}\frac{G M_p \mu}{k_B T},
\end{equation}
where $\gamma$ and $\mu$ are the gas adiabatic index and mean molecular weight respectively, and $k_B$ is Boltzmann's constant.

From this we can calculate a rainout time
\begin{equation}\label{eq:u-rain}
    t_\mathrm{rain}=\frac{U_\mathrm{rain}}{L_\mathrm{rain}}
\end{equation}
The values of the model parameters appear in Table~\ref{tab:pars}.

\begin{table}[]
    \centering
    \begin{tabular}{l l l }
     parameter & symbol  &  value\\
     \hline
      gas adiabatic index & $\gamma$ & 1.45 \\
      mean molecular weight & $\mu$ & 2.35 \\
      silicate evaporation temperature & T$_\mathrm{vap}$ & 2500\,K \\
      radiative opacity & $\kappa$ & 0.1\,cm$^2$/g\\
      latent heat & u$_\mathrm{lat}$ & 1.5$\times 10^{11}$erg/g \\
      core density & $\rho_c$ & 5\,g/cm$^3$ \\
      initial core mass & $M_c$ & 2\me \\
       \hline
    \end{tabular}
    \caption{Values of parameters used in the analytical calculations in section~\ref{sec:anal}, based on \cite{brouwers20} and \cite{ormel21}. In the numerical simulations these parameters vary with pressure and temperature.}
    \label{tab:pars}
\end{table}

\subsection{Timescales of interior evolution}\label{sec:times}

To give intuition on the thermal evolution of non-uniform polluted envelopes, we provide order of magnitude estimates of relevant timescales:

\underline{Settling}: 
For {silicate} vapor condensation we assume rain droplets sizes of $a_d=0.01-1\,\mathrm{mm}$, which provides, based on \cite{movsh08} 
sedimentation velocity of $v_\mathrm{sed}=10^{11}\times a_d^2\,[m/s]$, resulting in sedimentation time ($\tau_{sed}=R_p/v_{sed}$) of up to 200 yr, orders of magnitude {below} the Kelvin-Helmholtz times in the evolution model. 
Thus, taking settling to be an instantaneous process in our simulation is reasonable. 

\underline{Mass loss}: 
Average mass loss rate by XUV radiation from planets at 0.2\,AU is about $10^{-3}-10^{-1}$\me\ Myr$^{-1}$, using standard photoevaporation model \citep[e.g.,][]{owenwu13}. Estimated mass loss timescale of $\tau_{ML}=M_\mathrm{env}/\dot{M}$ results in timescale of $10^8-10^{10}$\,yr, {indicating that polluted planets in the lower mass} range can lose their entire envelopes. 

\underline{Conduction}:
Suppression of convection in the deep interior can form thermal boundaries in which heat is transported by conduction / layered-convection, where conductive transport is slower and therefore the lower bound for heat transport. The timescale for conduction in layer of thickness $X$, density $\rho$, and thermal conductivity $K$ is approximately $\tau_{cond}=X^2 c_p \rho/K$. For heat capacity $c_p=1\,[KJ/kg/K]$ and values for density and thermal conductivity from our simulations we get $\tau_{cond}=10^{-2}\times (X[m])^2\,[yr]$. Thus, an effective thermal boundary (i.e., $\tau_{cond}>10^9\,yr$) has a conductive layer of about 300\,km in order to keep most of the heat from formation in the deep interior in Gyr time. 

\underline{Material diffusion}: 
Diffusion in composition boundaries can be approximate by a Brownian motion with timescale $\tau_{diff}=L^2/2D$, where $L$ the length scale of the change in particle position, and $D$ diffusivity.
Estimate of diffusivity by Stokes-Einstein equation $D=\frac{k_B T}{6\pi \eta a}$, for temperature $T=5\times10^3\,[K]$, viscosity of $\eta=1\,[Pas]$, and {silicate} ($SiO_2$) molecular size of $a=10^{-9}\,[m]$, provides $D\sim 10^{-12}\,[m^2/s]$. Hence, diffusion can {transport} particles within $\tau_{diff}=1$Gyr to about 0.5\,km. 
Since this upper bound of diffusion length-scale is smaller than the conductive boundary, diffusion is expected to be negligible in the interior evolution of rock dominated planets.  
Higher viscosity (than 1\,Pas) and/or larger particles (than $10^{-9}$m) result in even slower diffusion. 

\section{Results}

\subsection{From formation to evolution}

The evolution calculation starts at disk dissipation. We find that the decrease in pressure of the outermost envelope at disk dissipation does not lead to mass loss by Roche lobe overflow, unless atmospheric metallicity (opacity) is unrealistically high.
If the upper atmosphere maintains a high opacity {for} some reason this would lead to envelope expansion and to mass loss \citep{ginzburg16}. However, the efficient grain settling expected in planetary atmospheres \citep{brouwers21,movsh08,ormel14,mordasini14} leads to grain-free atmospheres, thus low atmospheric metallicity. Using the grain-free opacity tables of \cite{freedman14} results in efficient cooling and no mass loss at the disk dissipation stage (see also appendix A.2 in \citealp{vazan18c}). 
This finding holds for all models we {have} tested in this work. 

\begin{figure}
\centerline{\includegraphics[width=9cm]{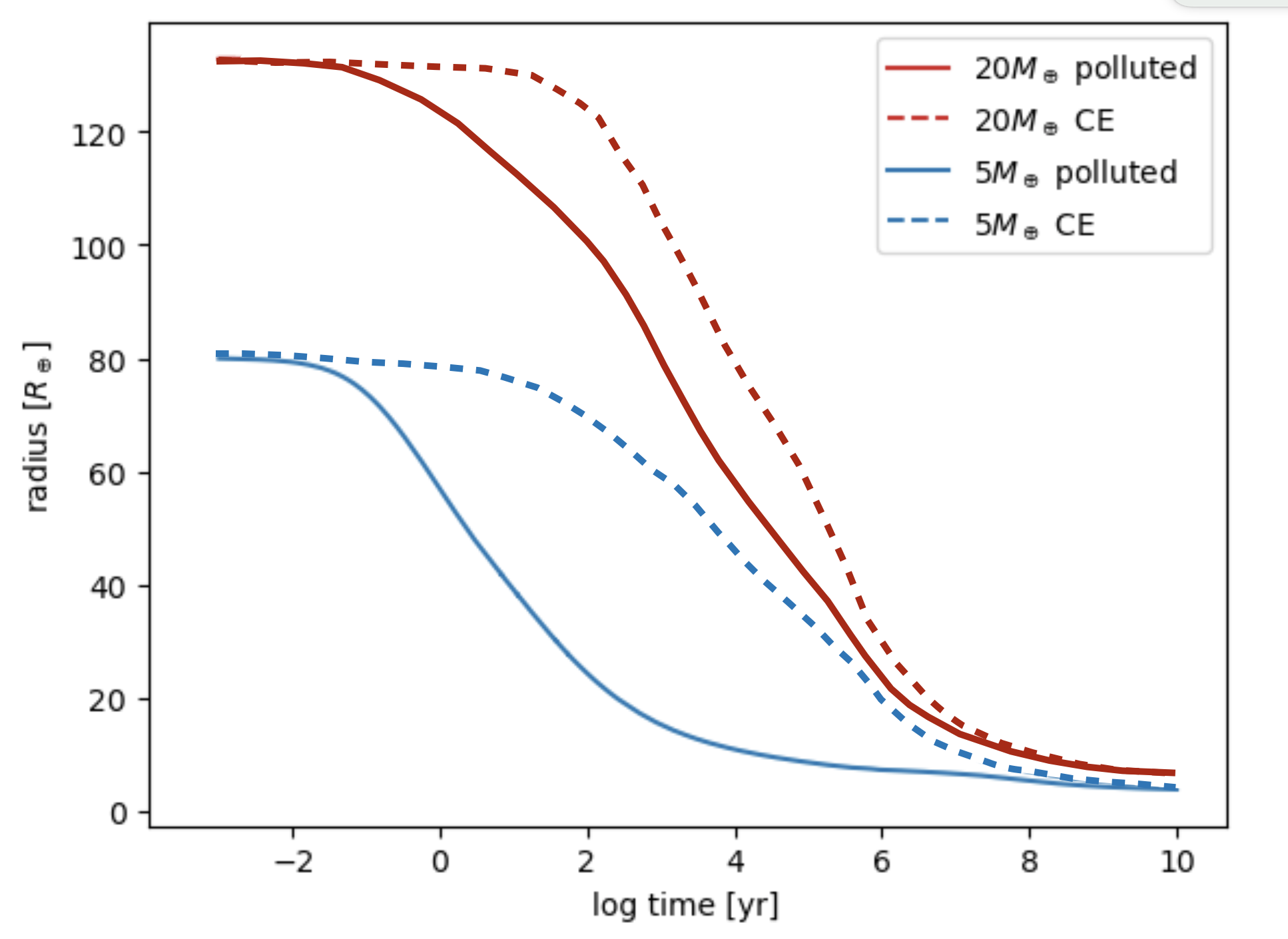}}
\caption{Radius evolution from disk dissipation of planets with polluted envelope (solid) and planets with same composition and core-envelope (CE) structure (dashed). Planets have static interior structure (no rainout) and are located at 0.2\,AU from a sun-like star. 
} \label{fig:Rev_early}
\end{figure}

After the disk dissipates the young planet enters a phase of rapid contraction of the outer diluted envelope, in which the formed planet contracts from its Bondi / Hill radius to about a tenth of it. 
In figure~\ref{fig:Rev_early} we show the radius contraction of the planets from figure ~\ref{fig:form} at 0.2\,AU. For comparison, we show planets with the same composition that formed with core-envelope structure. The initial models for the core-envelope cases are of isentropic core and envelope, with same central temperature as in our formation models. 
Evolution time is shown in log scale to emphasis the early evolution right after planet formation ends. 
In general, the early rapid contraction of planets with polluted envelopes is faster than that of similar planets with core-envelope structure. The higher mean molecular weight of the polluted envelope and the slower heat transport due to the deep composition gradient (convection suppression) lead to accelerated early contraction of the polluted envelopes\footnote{This finding is for static interior structure. When silicate rainout takes place the radii of planets with polluted envelopes exceed that of planets with core-envelope structure, as will be discussed in the next sections.}, where the difference is greater for smaller, more-polluted, planets. 
Timescale for the contraction from formation radius to about tenth of it varies from a few $10^4$ years for the 5\me planets to a few $10^7$ years for 20\me planets. 
Higher opacity could potentially {prolong} the contraction time, but we find the required opacity increase unrealistically high.

\subsection{Thermal and structural evolution}

After the rapid contraction that heats-up the outer envelope, the planet starts to cool. The cooling of the uniform metal-rich envelope is governed by large scale convection, while the heat transport from the deeper interior - where the composition gradient acts as a thermal boundary - is less efficient. 
Consequently, the temperatures in the metal-rich envelope decreases to below the saturation level and silicate condensate and settle (rainout) to deeper undersaturated layer (see figure~\ref{fig:rain}).
The process of rainout is a top-to-bottom process, in which a complete rainout leads to a core-envelope structure. 

Core erosion and mixing upwards of the silicates by convection (convective-mixing) are found to be negligible in all the planets we explored in this work, for two reasons; first, the deep composition gradients from formation are relatively steep in metal-dominated planets, and therefore are stable against convection \citep{vazan22}; and second, inhibition of convection by composition gradients along the saturation pressure-temperature profile suppresses redistribution of elements in saturated layers \citep{guillot95a,markham22}.

\begin{figure*}
\centerline{\includegraphics[width=\textwidth]{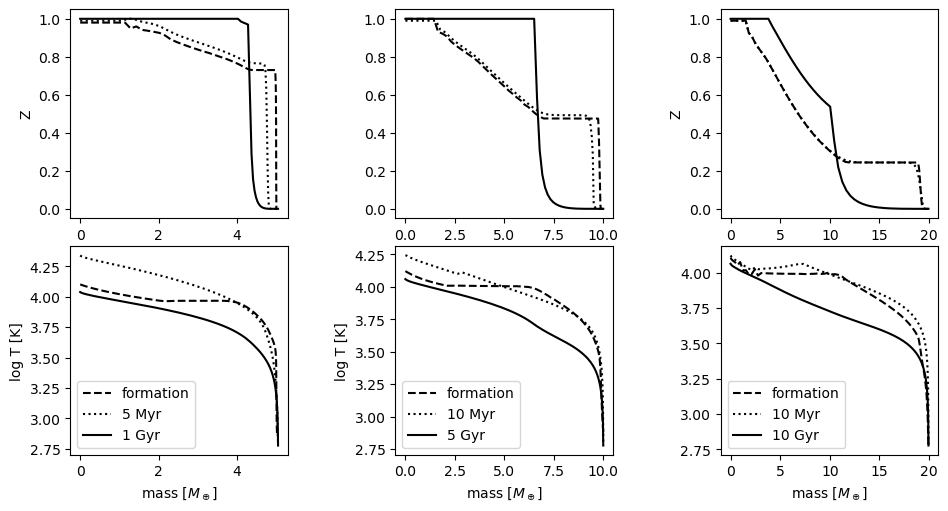}}
\caption{Interior silicate mass fraction (top) and temperature profile (bottom) of 5\me (left), 10\me (middle) and 20\me (right) planets. Profiles are shown at the end of formation (dashed), after the rapid contraction (dotted) and after 1, 5, 10 Gyrs (solid) respectively, for evolution at 0.2\,AU. The early contraction (from Hill sphere to about one tenth of it) is the cause of the gas temperature increase between formation and the Myr {curves}. 
{The increase in central temperatures is the result of core contraction, due to the constant core density assumption in the formation model.} 
The dashed and dotted lines of the 20\me model overlap {in the top-right panel.}}\label{fig:ZTm}
\end{figure*}

In figure~\ref{fig:ZTm} we show silicate mass fraction (Z) profile and temperature profile of  5\me (left), 10\me (middle), and 20\me (right) planets located at 0.2\,AU. {It shows that} the early rapid contraction phase has only a small effect on the composition distribution (dotted). 
Despite the large change in radius, the deep interior (composition gradient and below) stays almost unchanged in this early evolution phase. 
In the long term cooling {phase} the composition distribution significantly changes by rainout. Rainout in the 10\me and 5\me planets results in a core-envelope structure before 5\,Gyr and 1\,Gyr, respectively. In the more massive and gas-rich 20\me planet rainout depletes the outer layers from silicates but yet a significant composition gradient is maintained in the deep interior {after even} 10 Gyr. 

\begin{figure*}
\subfigure{\includegraphics[width=0.332\textwidth]{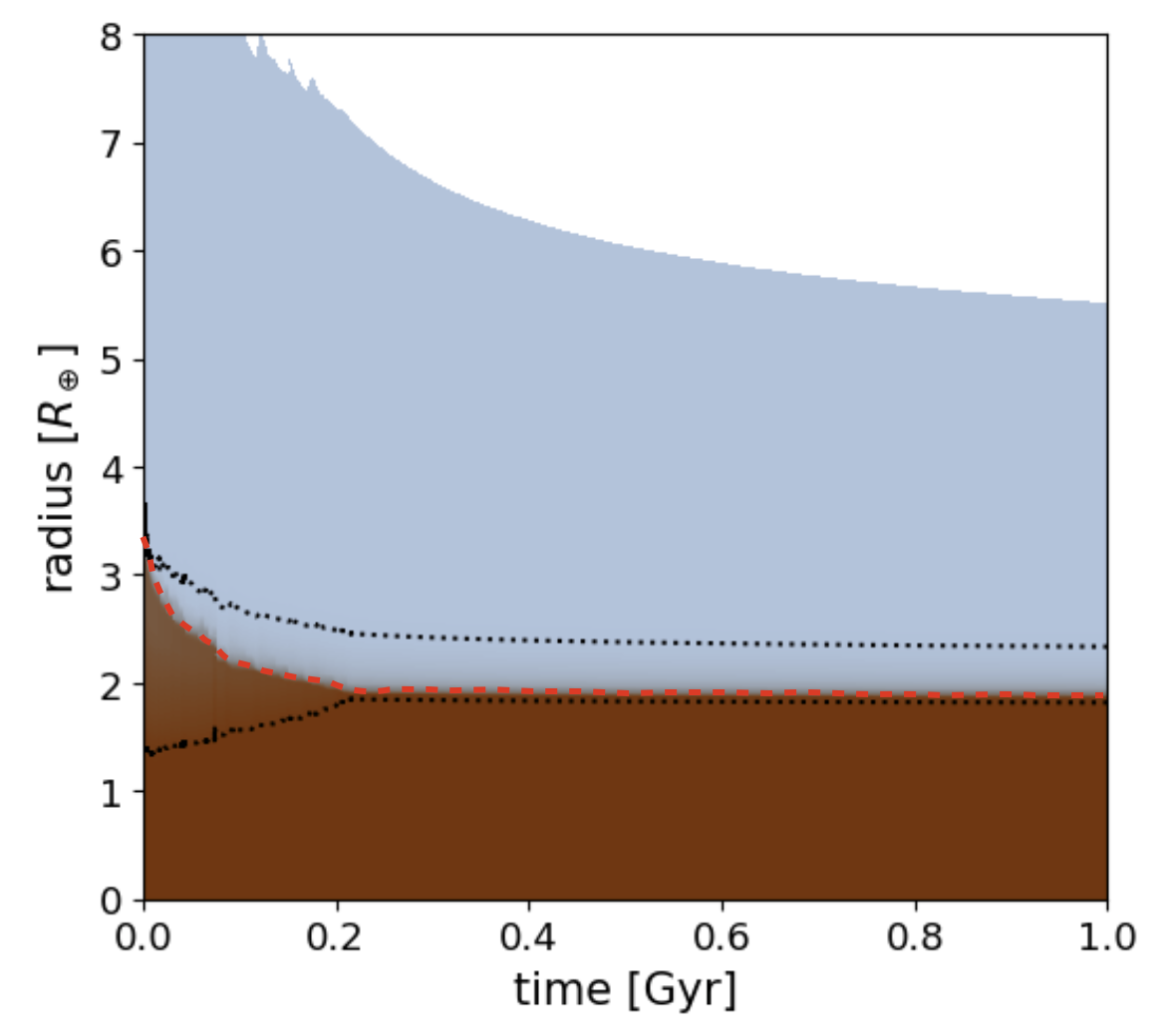}}
\subfigure{\includegraphics[width=0.310\textwidth]{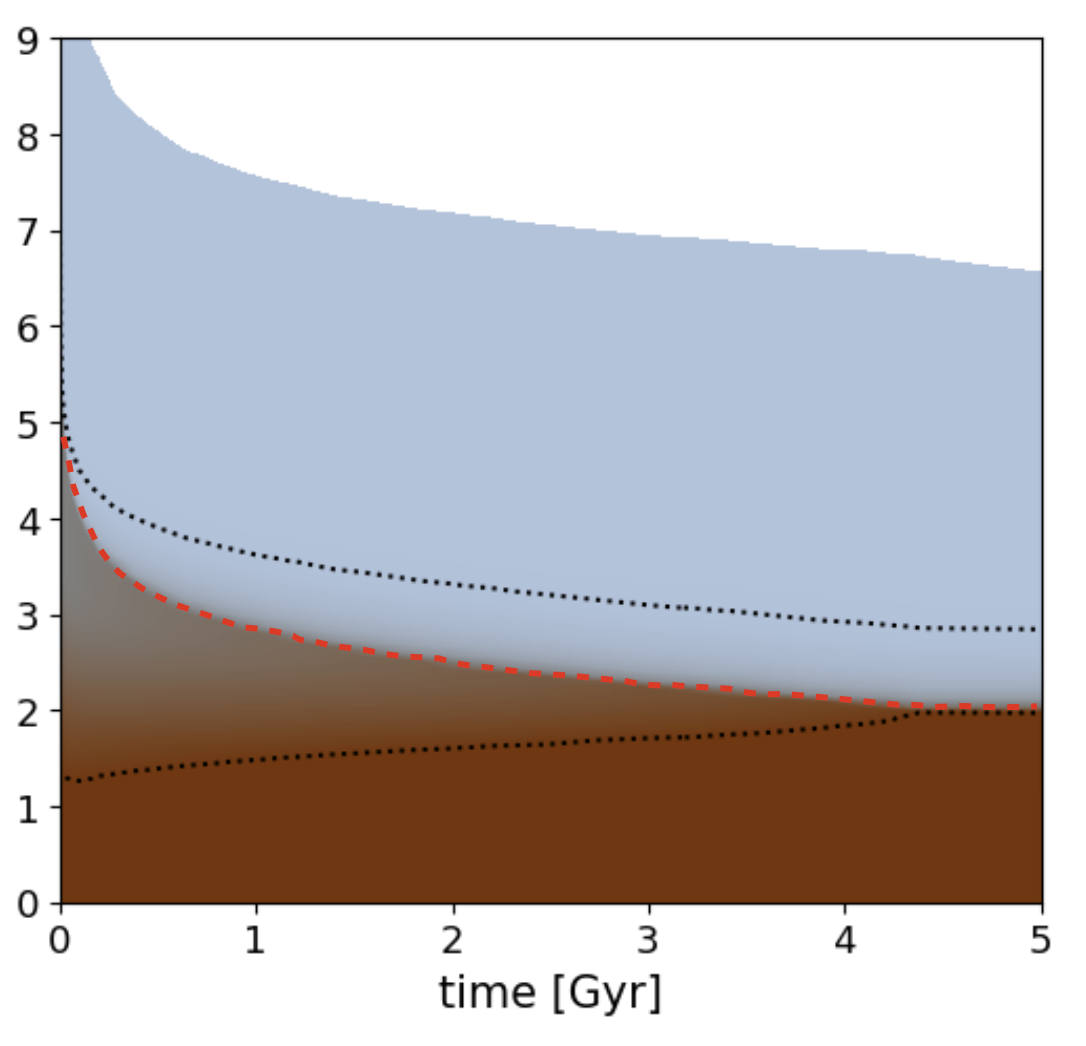}}
\subfigure{\includegraphics[width=0.358\textwidth]{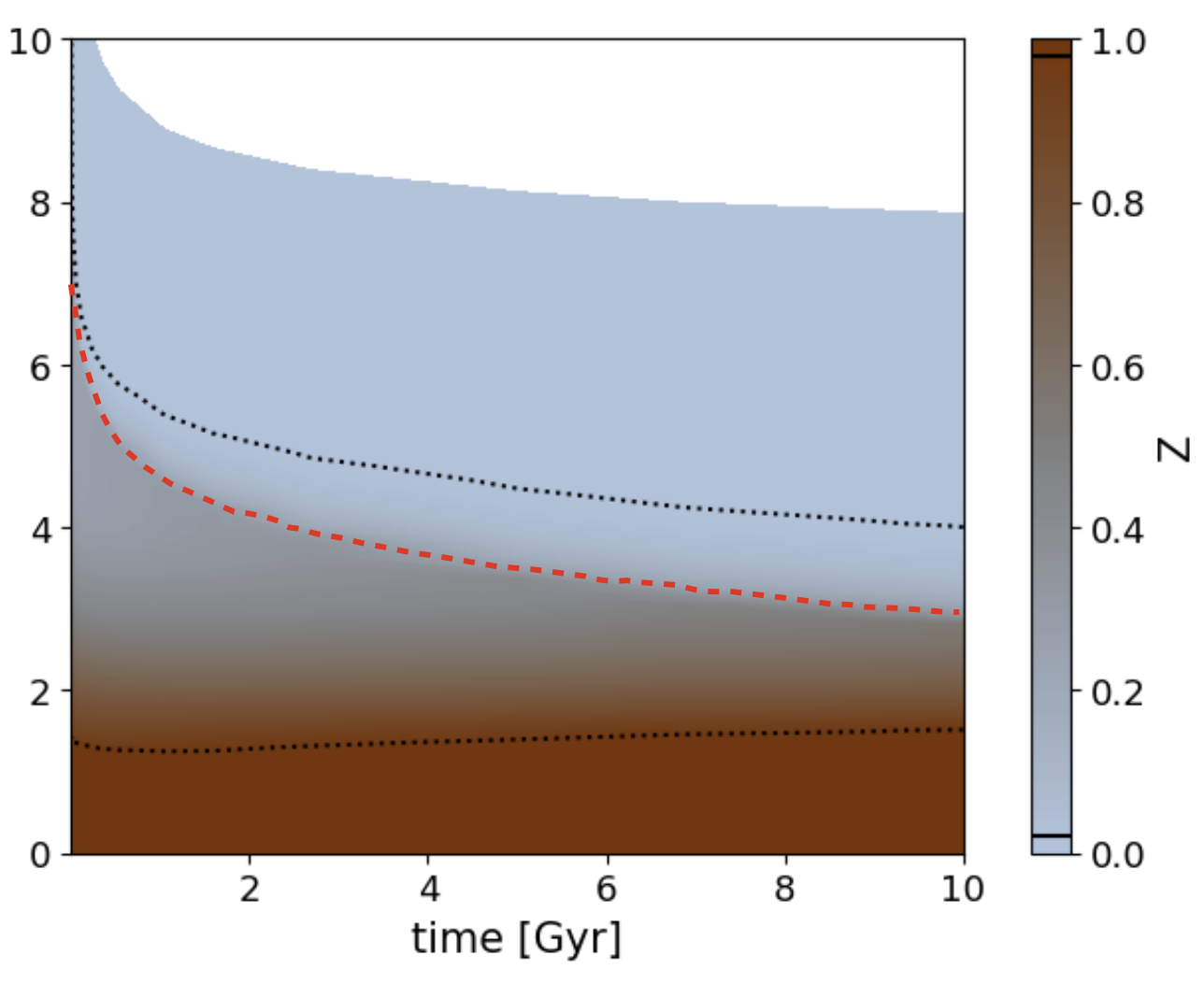}}
\caption{Silicate mass fraction (color) distribution in radius (y-axis) and time (x-axis). Evolution is shown for 5\me (left), 10\me (middle) and 20\me (right). 
Notice the different ranges for the x-axis (1\,Gyr, 5\,Gyr, 10\,Gyr, respectively). Black dotted curves indicate Z=0.98 and Z=0.02 silicate mass fraction levels. Dashed red curve signified the boundary between saturated and undersaturated regions, according to equation~\ref{eq:sat}.
{A core-envelope boundary is approximately indicated by the $Z=0.98$ curve. Because of the high temperatures at the core-envelope boundary, the (saturated) envelope is characterized by a silicate mass fraction of a few percent.}}
\label{fig:Zrt}
\end{figure*}

The structure evolution (redistribution of silicate in time) by the rainout is shown in figure~\ref{fig:Zrt}. Rainout of silicate in the 5\me planet starts almost immediately after disk dissipation and ends (the planet reaches core-envelope structure) after 0.38\,Gyr. The process is much slower in the gas-rich 20\me planet, where silicate rainout starts after 10 Myr, and is still ongoing after 10 Gyr. The 10\me planet is an intermediate case, in which core-envelope structure is reached after 4.25\,Gyr.

\begin{figure}
\centerline{\includegraphics[width=9cm]{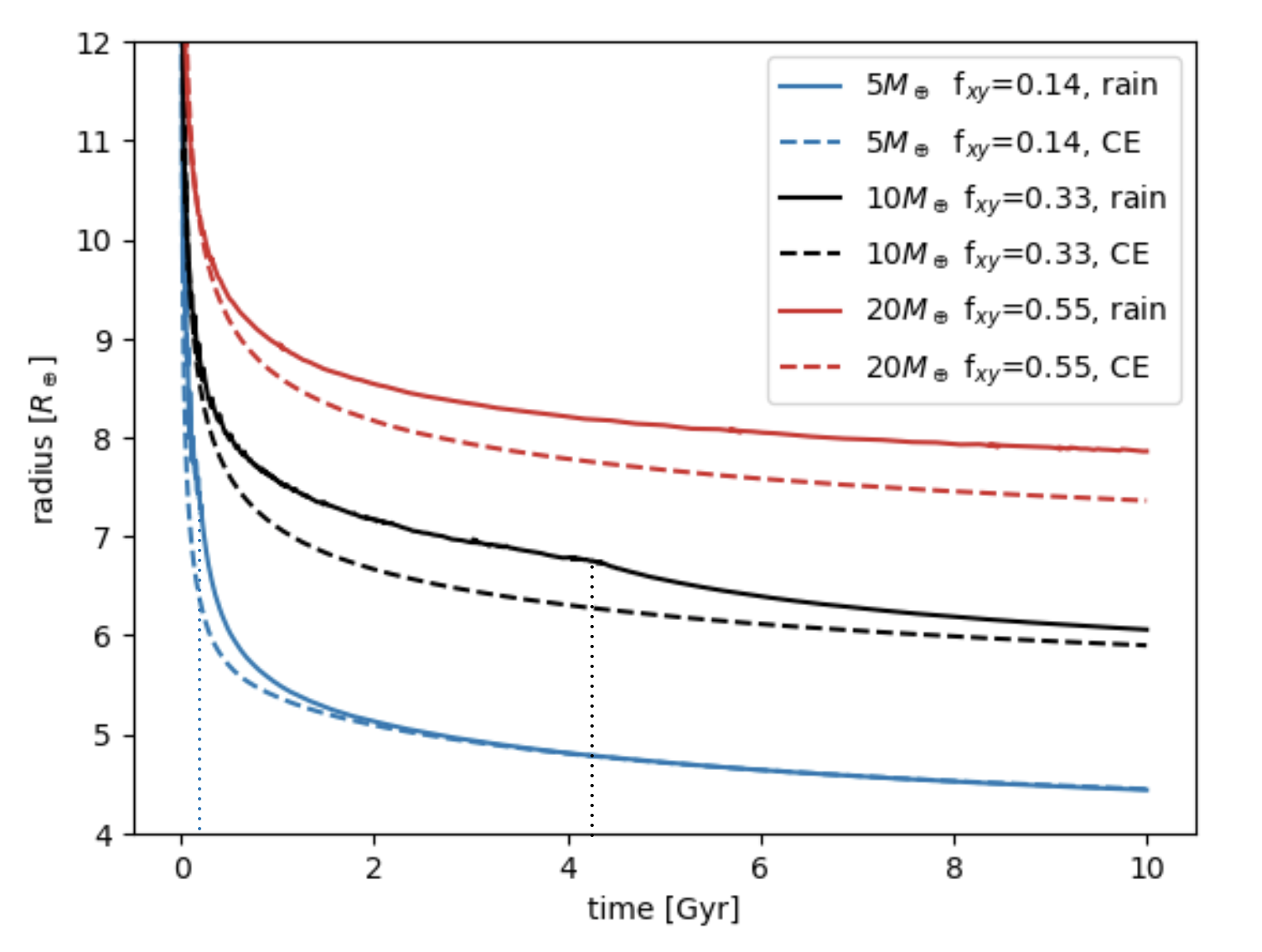}}
\caption{Radius evolution of planets formed with polluted envelope (solid) and planets with the same bulk composition but  {starting with a} core-envelope (CE) structure {from the outset} (dashed). {The measured rainout time is determined by the time where the classical core-envelope structure is established (dotted {vertical} line).} Rainout in the polluted envelopes causes radius inflation in comparison to planets formed with core-envelope. Radius inflation is shorter and {more prominent} in smaller planets. All planets located at 0.2\,AU from a sun-like star.}\label{fig:Rev}
\end{figure}

\subsection{Rainout dependency on envelope mass and metal mass}

Planets with different envelope masses experience rainout at different time and {strength} \citep{vazanormel23}. Consequently, the strength and duration of radius inflation, caused by the rainout energy release, varies; when rainout is faster, the energy release is on a shorter timescale and therefore radius expansion is larger. Rainout is faster and earlier in smaller planets, due to their lower envelope mass, larger envelope metallicity, and lower gravity. In figure~\ref{fig:Rev} we show the radius evolution {corresponding to} the 5\me, 10\me and 20\me planets shown in figure~\ref{fig:Zrt}. For comparison we also plot the radius evolution of planets {of} the same composition {but that start off} with a core-envelope structure {from the outset}. 
As can be seen, the 5\me planet has an early and large radius inflation by rainout, after which it becomes identical to its core-envelope twin. The 10\me and 20\me keep a moderate radius inflation in comparison to their core-envelope analogue, due to post-rainout heat release (10\me) and ongoing rainout (20\me). 

\cite{misener22} calculated structure model for planets with polluted envelopes and found that silicate vapor layer decreases the planet’s total radius compared to a similar planet with pure H,He envelope. While mean molecular weight and composition gradient (suppression of convection) arguments indeed lead to smaller radius as suggested by these authors, including the structure evolution and specifically the rainout process in thermal evolution can result in larger radius, caused by the release of rainout energy: latent heat, gravitational energy, and late release of locked formation energy. 

The envelope mass, more specifically the pressure-temperature at the bottom of the envelope, plays a key role for the rainout time. Planets with light envelopes experience shorter and earlier rainout than planets with massive envelopes. In \cite{vazanormel23} we found that an envelope mass of approximately 0.75\me to be an upper limit for sub-Neptunes to reach the core-envelope structure within 1 Gyr. This result was found for planets with 6.7\me of silicates. 
Here we repeat this calculation for planets with different silicate masses.

In figure~\ref{fig:train_comp} we show simulation results (points) of rainout timescale with gas (H,He) mass for three different cases of total silicate mass ($M_Z$): 4.3\me (blue), 6.7\me (red), 8.9\me (green). The lines are for analytical fits and will be discussed in the next section. 
We find that the rainout timescale increases with the total silicate mass (for same envelope mass), thus rainout takes longer in more massive planets with the same gas mass.  
Simulations results indicate that the change in rainout time with silicate mass is especially noticeable in planets with small gas to silicate ratio. For example, rainout in a gas envelope of 1.2\me H,He takes 0.81\,Gyr in 5.5\me planet, 1.23\,Gyr in $\sim$8\me planet, and 2.1\,Gyr in $\sim$10\me planet. 
The reason is that stronger gravity in more massive planet leads to denser and hotter deep envelopes, in which silicate are undersaturated for longer. 
However, when gas mass is comparable to the metal mass the differences between cases diminishes, probably the result of the larger heat capacity of hydrogen and helium in comparison to rock. 

\subsection{Comparison to analytical calculation}\label{sec:comp}

We next use our detailed simulation to examine the validity of the analytical results of \cite{brouwers20} on the rainout timescale. As described in section~\ref{sec:anal} the analytical model contains assumptions on the structure (three layer model) and its properties (ideal gas, constant parameters), and is obtained by assuming low mass envelopes (to neglect gas gravity). 
The free parameter values were set as in \cite{brouwers20}, and appear in table~\ref{tab:pars}. 
The fit of the analytical calculation (equation~\ref{eq:u-rain}) to the simulation results is found to be poor for planets with a significant gas fraction. The poor fit may be the result of neglect of envelope gravity, oversimplification of gas compression (ideal gas), and constants such as opacity, core density, and evaporation temperature. 
However, we obtain a much better fit to the simulation data when we take:
\begin{equation}\label{eq:train}
    t_\mathrm{rain} = \frac{U_\mathrm{rain}}{L_\mathrm{rain}} \times 0.0251\left(1+\frac{M_{xy}}{M_\mathrm{z}} \right)^{5.2}
\end{equation}

\begin{figure}
\subfigure{\includegraphics[width=9cm]{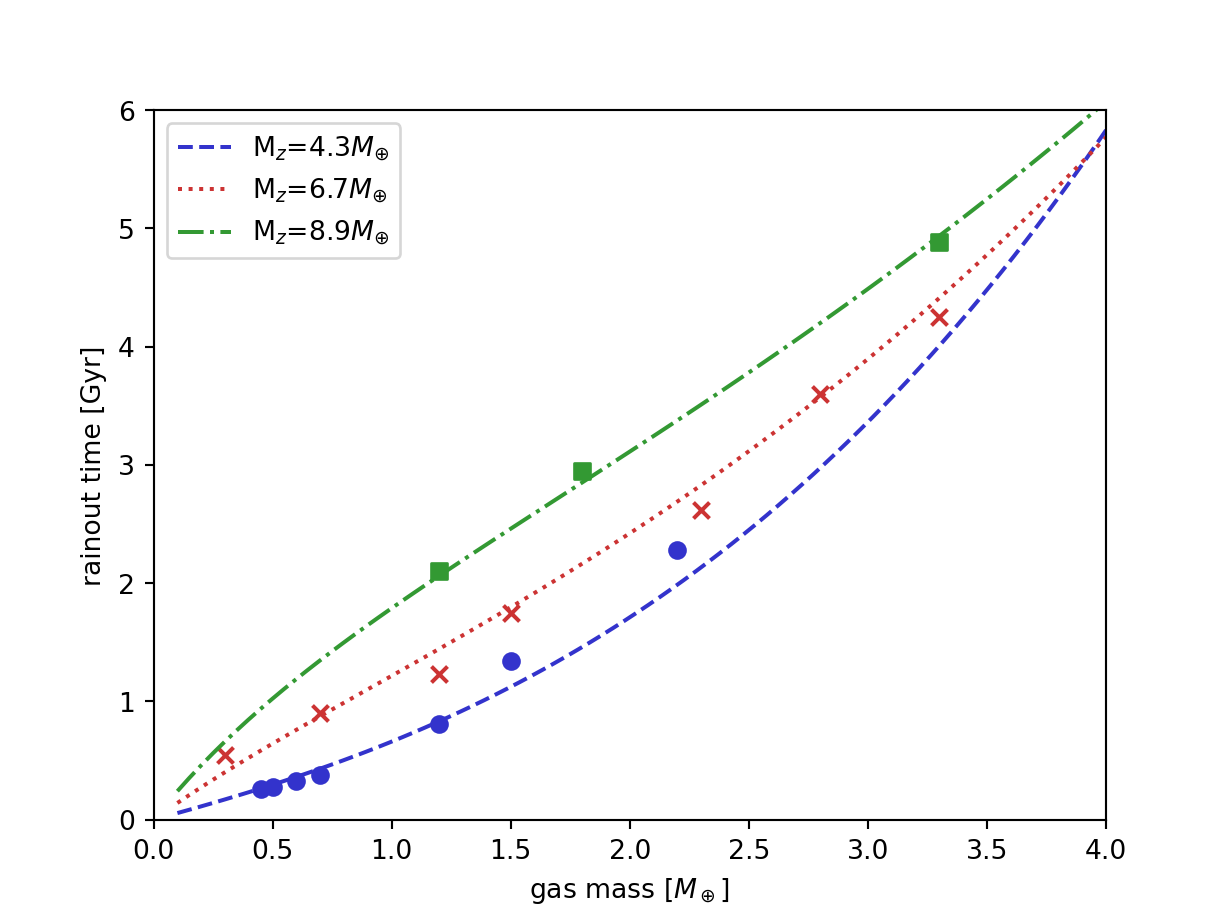}}
\caption{Rainout time as a function of gas (H,He) mass for three cases of silicate mass: $M_Z$=4.3\me (blue), $M_Z$=6.7\me (red) and $M_Z$=8.9\me (green). Shown are numerical simulations (points) and analytical calculation (curves).   
Analytical model is according to equation~\ref{eq:train} and parameters appear in table~\ref{tab:pars}.}\label{fig:train_comp}
\end{figure}

The improved fit of equation~\ref{eq:train} to the simulations is achieved by the additional term of gas to silicate mass ratio, which leads to a steeper dependency of rainout timescale in envelope mass $M_{xy}$, in comparison to the old calculation (equation~\ref{eq:u-rain}). 
The inverse dependency in silicate mass is less significant than the envelope mass, because $M_\mathrm{z}$ appears also in the numerator of equation~\ref{eq:u-rain}. Thus, the rainout timescale increases steeply with envelope mass, and only moderately with metal mass, as found in the simulations. 
In figure~\ref{fig:train_comp} we also show the rainout time vs. planet's gas mass calculated by the analytical approximation in equation~\ref{eq:train} (curves) for the three cases of silicate mass from the numerical simulations. 

Important source of the difference between the simulation and the analytical calculation is the use of ideal gas in the analytical calculation, an approximation that suffers from overestimated compressibility and constant heat capacity. In appendix~\ref{sec:app1} we present a comparison between ideal and non-ideal (tabular) H,He EoS in formation-evolution models. 

Next, we derived an empirical approximate, based on our simulations, for a relation between the rainout time (the time from formation it takes to reach a core-envelope structure) and the radius inflation at the end of rainout time, which is the maximal radius inflation. 
We define the radius inflation $\Delta R$ as the {excess} radius {a planet undergoing rainout has compared to a planet of} similar mass, composition, age formed with a {classical} core-envelope structure.  
We find that a maximum radius inflation is approximately proportional to the inverse square-root of the rainout timescale:
\begin{equation}\label{eq:dRt}
    \left( \frac{\Delta R}{R} \right)_\mathrm{max} 
    \simeq A \left( \frac{t_\mathrm{rain}}{\mathrm{1\,Gyr}} \right)^{-0.5}
\end{equation}
For planets with silicate mass of 4.3\me a proportion factor of $A=0.14$ fits the simulation results. 

\begin{figure}  
\centering{\includegraphics[width=\columnwidth]{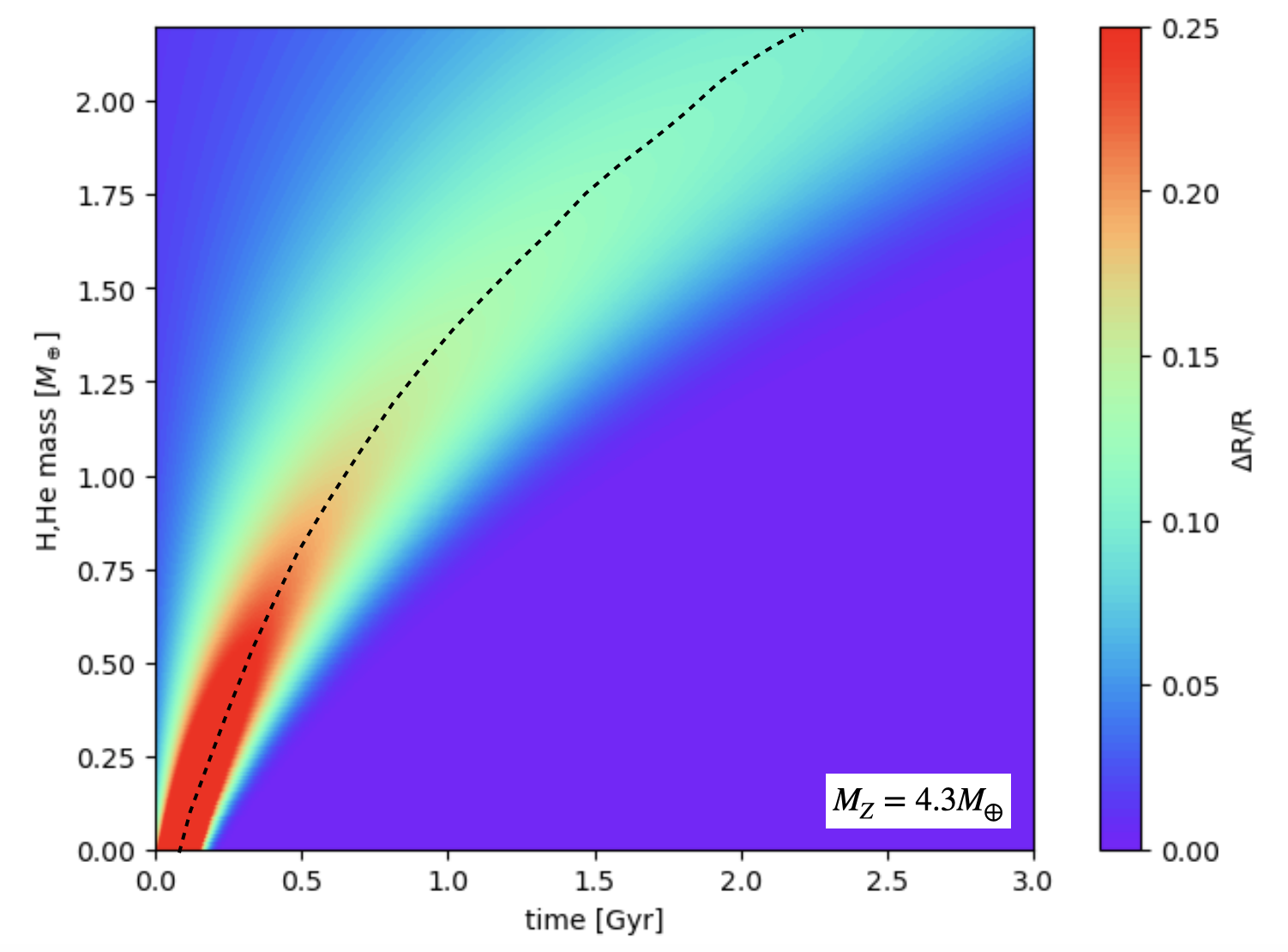}}
\caption{Radius inflation, $\Delta R/R = (R-R_{CE})/R_{CE}$, as a function of age (x-axis) and envelope mass (y-axis), for planets with 4.3\me of silicates. For {illustrative }clarity we limited the scale of $\Delta R/R$ to 0.25 (red color), but {in reality} $\Delta R/R$ {reaches values} up to 1. Dashed curve signifies the maximum radius inflation for each envelope mass.}\label{fig:dR3D}
\end{figure}

We then use equations~\ref{eq:train} and~\ref{eq:dRt} to obtain the maximum normalized radius inflation as a function of envelope mass in time (age). 
The radius inflation difference between a raining-out planet (R) and a core-envelope planet (R$_{CE}$) with same composition ($\frac{\Delta R}{R}=\frac{R-R_{CE}}{R_{CE}}$) is time dependent. 
For the time dependency we fit the simulation results with a Gaussian, in which the standard deviation is the inverse of maximum radius inflation, and the mean is the rainout timescale. 
The results ($\Delta R$/R) for planets with 4.3\me of silicates are presented in figure~\ref{fig:dR3D}. 
As is shown, radius inflation is larger and earlier for planets with lighter envelopes. 

\subsection{Gas escape enhancement by rainout:  Sedimentation-powered mass loss}\label{sec:ML}

In the case of a static (non-evolving) interior, the polluted envelopes may diminish mass loss by photoevaporation, due to higher envelope density and smaller radius. However, in an evolving interior the energy release by the rainout processes enhances mass loss, due to radius inflation.  
{Mass-loss by photoevaporation is usually modeled by the energy-limited approach \citep[e.g.][]{lammer03,owenwu13,lopezfor13}, where planetary evaporation depends on the radius to the third power. From \cite{rogersowen21}:
\begin{equation}\label{eq:ml}
    \dot{M} = \eta \frac{\pi R_p^3 L_{\rm XUV}}{4\pi d^2 G M_p}
\end{equation}
where $\eta = 0.17 (v_{\rm esc} / 25\,[km/s])^{-0.4}$ is efficiency coefficient, depending on the the escape velocity ($v_{\rm esc}$), and XUV luminosity $L_{\rm XUV}$ from a Sun-like star is
\begin{equation}
    L_{\rm XUV} = 
\begin{cases}
    10^{-3.5} L_{\odot} & \text{for  } t < 100 Myr \\
    10^{-3.5} L_{\odot} \left( \frac{t}{100\,Myr}\right)^{-1.5} & \text{for  } t \geq 100\,Myr
\end{cases}
\end{equation}
}

According to the equations above, larger radius, as a result of radius inflation by rainout, is expected to significantly increase mass loss rate, especially when the inflation takes place early on. 
But the other way around also holds: the loss of hydrogen from the surface of planets with polluted envelopes accelerates rainout via two channels - it raises the silicate to gas ratio in the outer envelope thus increasing oversaturation, and it removes part of the gas "blanket" thus hasten the cooling. 
Therefore, if rainout timescale overlaps with the period of strong stellar XUV radiation the mass loss by photoevaporation, which is proportional to radius cube (eq.~\ref{eq:ml}) grows significantly, boosting more radius inflation, which in turn enhances mass loss.
Consequently, close-in and less massive planets with polluted envelopes are more vulnerable to enhanced mass loss by photoevaporation. 

\begin{figure}  
\centering{\includegraphics[width=9cm]{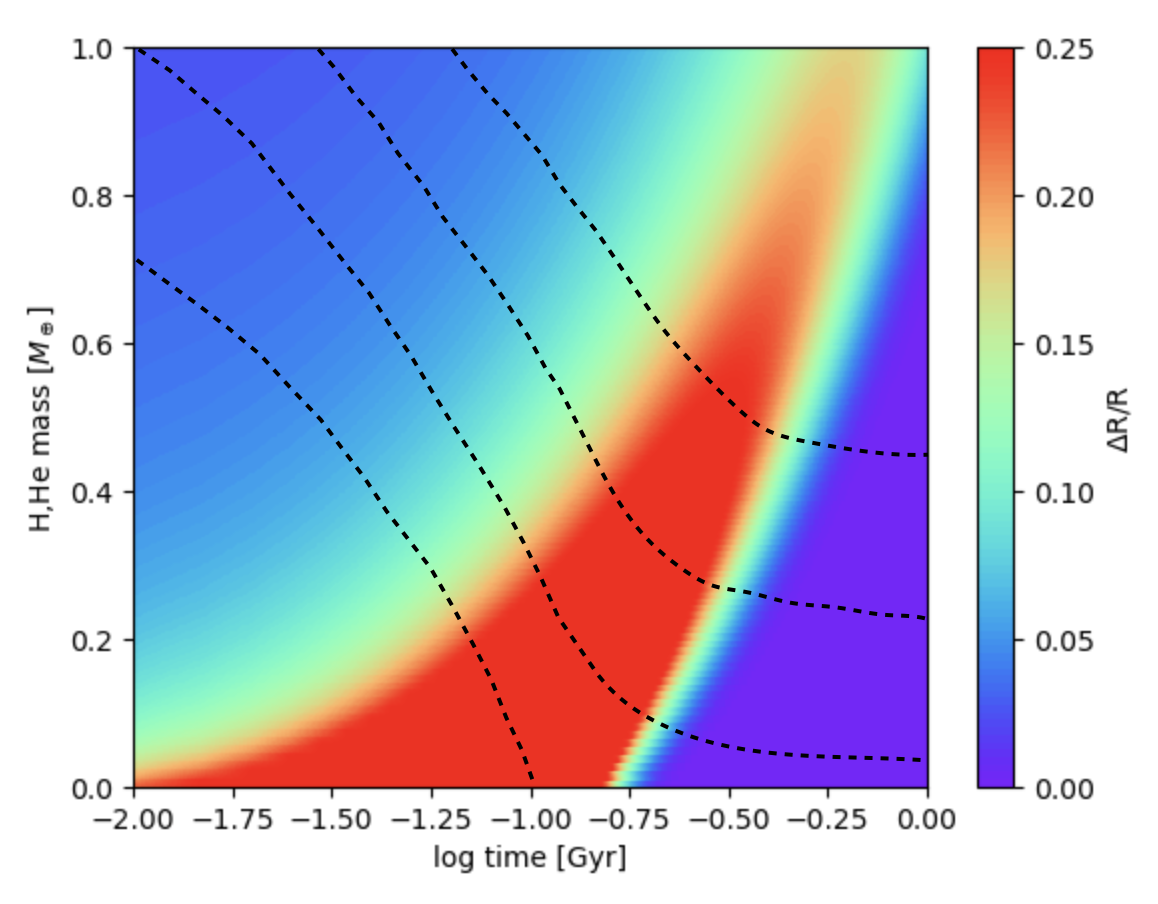}}
    \caption{Zoom-in on Fig.~\ref{fig:dR3D} in the first 1\,Gyr, in log time scale. Planets with significant radius inflation in the first 0.1\,Gyr are vulnerable to mass loss by photoevaporation. As in figure~\ref{fig:dR3D}, red color signifies $\Delta R/R$ in the range of 0.25-1. Dashed curves show examples of estimated mass loss tracks at 0.2\,AU.}
    \label{fig:dR3D_1Gyr}
\end{figure}

Figure~\ref{fig:dR3D_1Gyr} is a zoom-in of the first 1\,Gyr of figure~\ref{fig:dR3D} in log time (x-axis). As can be seen, super-Earth planets with envelope masses below 0.4\me experience a major envelope expansion by rainout during the first hundreds of Myrs. This is when XUV-driven photoevaporation rates peak, then enhanced by "sedimentation-powered" mass loss. 
Radius inflation of raining polluted envelopes that is larger by 0.25 ($\Delta R/R=0.25$) doubles the mass loss rate, and larger radius inflation of $\Delta R/R=1$ is translated to eight times more mass loss in comparison to a similar planet with initial core-envelope structure. 
The enhanced "sedimentation-powered" mass loss, caused by the mutual effects of mass loss and rainout, may end in a complete strip of the gas envelope in planets that were born with polluted low-mass envelopes. 
A further investigation would require coupling the two processes to examine the mutual effect in detail. 

\subsection{Model parameter dependency}\label{sec:pars}

\begin{figure}
\subfigure{\includegraphics[width=\columnwidth]{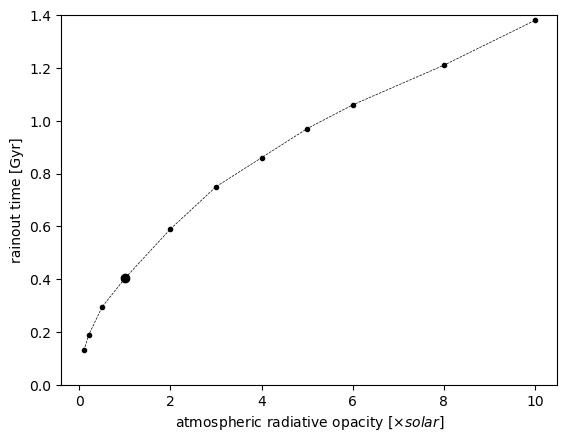}}
\caption{Rainout time as a function of atmospheric radiative opacity ($Z_\mathrm{min}$ in eq.~\ref{eq:Zeff}), based on the opacity formalism of \cite{freedman14}. Cases shown for 5\me planet with 4.3\me of silicates and 0.7\me gas. Points are simulations results, bigger point is the standard (solar metallicity) case.}\label{fig:train_op}
\end{figure}

Atmospheric metallicity has a major role in cooling of gaseous envelopes, {by changing the radiative opacity in the outer envelope,} acting as the bottleneck for the planet cooling. Consequently, it influences the silicate rainout timescale. The higher the atmospheric opacity the longer the rainout time. 
{Opacity can be affected by atmospheric phenomena like cloud formation or haze formation \citep{ormelmin19,poser24}.
{To explore the dependence of the atmospheric opacity on our results}, we modify the outer atmospheric radiative metallicity. {Specifically,} we obtain the radiative opacity change by adjusting the $Z_\mathrm{min}$ parameter in equation~\ref{eq:Zeff} {to values} between 0.1-10 times solar.}
As is shown in figure~\ref{fig:train_op}, changing the atmosphere metallicity by two orders of magnitude (between 0.1-10 times solar) changes the rainout time by an order of magnitude. 
{We remark that, although we have formulated the opacity in terms of a metallicity $Z_\mathrm{radiative}$, adjusting the opacity by changing the value of $Z_\mathrm{min}$ has no bearing on the actual metallicity (composition) and overall equation of state. In this way, adjusting $Z_\mathrm{min}$ will directly inform us how the opacity affects the rainout timescale.}

The effect of distance from the star on the results is not very strong and varies with mass. In massive planets and/or massive envelopes the distance from the parent star has negligible effect on an interior evolution and on the rainout process, as compressed gas density is determined mainly by gravity. 
In planets with light envelopes the irradiation effect can be noticeable. 
We find that the 5\me planet shown in figure~\ref{fig:Zrt} experiences 10\% shorter rainout phase at 1\,AU than at 0.2\,AU. This comparison is for the same planetary mass and composition. Indirect effect of irradiation via mass loss is ignored in this comparison (see section~\ref{sec:ML} for mass loss effects). 

We next examine the effect of planet formation conditions, such as pebble accretion rate and pebble size on the results. Planet formation conditions shape the initial temperature profile and metal distribution, yet varying the formation parameters within the range of \cite{ormel21} model has only small effect on the long term evolution. The reason is that planet formation parameters, unlike opacity and distance from the star, don't influence the long-term cooling rate very much.  

\subsection{Overview and implications on observation interpretation}\label{sec:over}

\begin{figure*}
\centerline{\includegraphics[width=18cm]{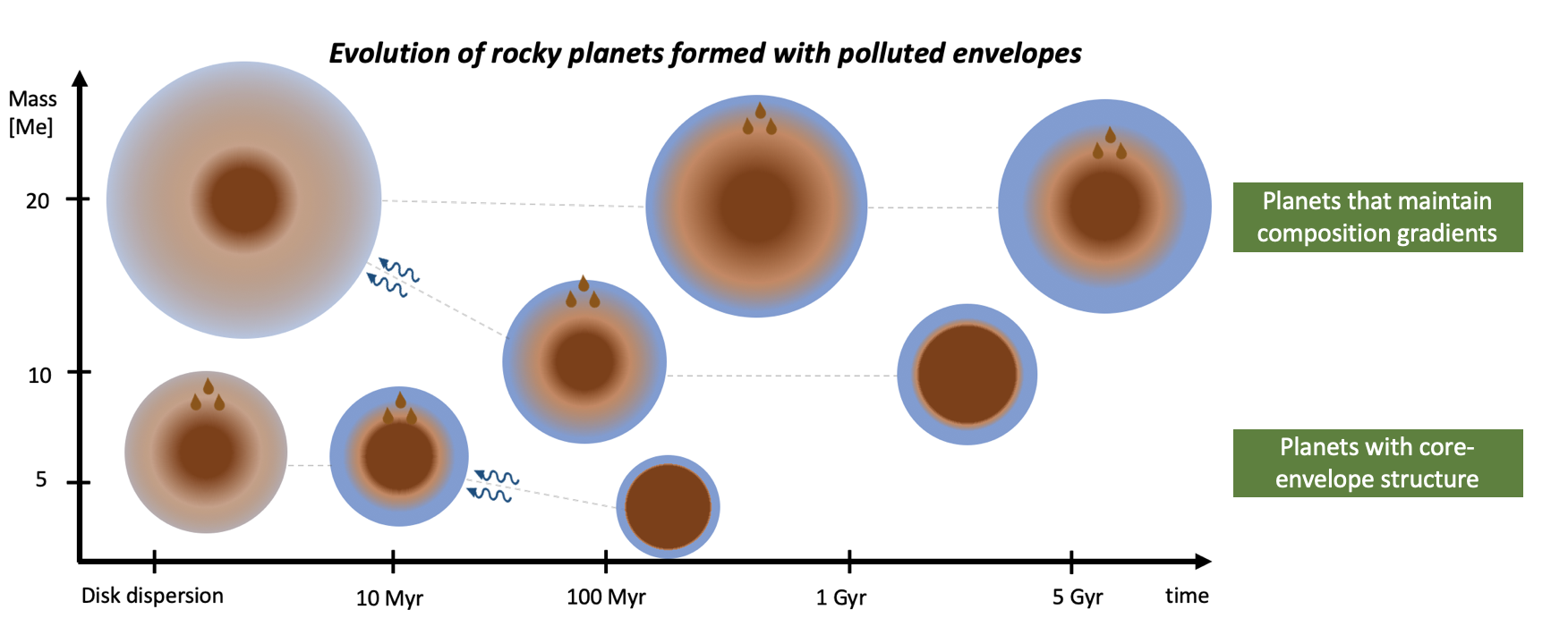}}
\caption{Summary sketch: interior evolution (x-axis) for different planetary masses (y-axis) formed with polluted envelopes. Planets can be classified into three groups regarding
their interior structure: planets with massive gas envelopes that maintain part/all of their composition gradients, planets with low-mass envelopes that completed their rainout and have core-envelope structure, and planets that lost all of their gas (not shown). See section~\ref{sec:over} for details. Phases of rainout and mass loss are marked in droplets and wave-arrows symbols respectively. Values in this schematic picture can vary with additional parameters such as atmospheric opacity and envelope to core mass.}\label{fig:cartoon_all}
\end{figure*}

Our results indicate that rocky super-Earth-to-Neptune mass exoplanets can be {classified into three groups regarding their} interior structures, depending mainly on their mass and in particular their gas (H,He) mass: 
\begin{enumerate}
    \item {Planets that have} lost their H,He envelope due to  "sedimentation-powered" mass loss and/or other mass loss mechanisms; 
    \item Planets that retain a few percent of gas (by mass) {after the mass-loss stage, whose interiors} have {since evolved to} a core-envelope structure; 
    \item Planets with thick envelopes that still have their interiors {characterized by a} composition gradient(s). 
\end{enumerate}
In figure~\ref{fig:cartoon_all} we show a schematic picture of the thermal and structural evolution of the planets that retain (part of) their primordial envelopes. The time and mass axes are for our standard set of parameters, and can vary with parameters as described in section~\ref{sec:pars}. 

The evolution of planets with polluted envelopes and the {subsequent} radius inflation by the rainout provides an alternative for the interpretation of {planets'} radius-mass relation. The observed mass-radius relation if the planet is raining-out is overestimated in the standard model (see \citealt{vazanormel23} for exoplanet examples). The main parameter to {determine whether} rainout takes place in a{n observed} planet is its age. Although currently many of the observed planets don't have good age estimates, in a few years {a much} larger and {precise} database of stellar ages {will be available due to} the PLATO mission \citep{rauer14short}, which will allow us to examine {the temporal dimension} of {the mass-radius relationship}. 

The heat flux {generated by} the interior of {a planet experiencing} rainout is {significantly} larger than in the standard (non-polluted) case, due to the {release of rainout energy} (potential energy, latent heat and formation energy). Certain chemical abundances, like the C/H ratio in the visible atmosphere {of a planet} may {provide clues about} the cooling history of the planet and its effective temperature \citep{fortney20}. Observations by JWST \citep{greene16jwst} and ARIEL \citep{tinetti18short} are expected to obtain spectra of a wide range of planetary atmospheres, and will help us identify {those} planets that formed with polluted envelopes.

The radius inflation caused by the rainout from polluted envelopes can explain, under certain conditions, the puffiness of observed close-in young planets, like the 0.5\,Gyr old Kepler-51 system \citep{masuda14,libby-roberts20}. In this scenario, mass loss and rainout mutual effect {result in} significant radius inflation that peaks at the observed age. The conditions to preform such evolutionary tracks require relatively high atmospheric metallicity and significant initial gas content. Moreover, if rainout is the dominant mechanism for puffing close-in young planets, then the current estimate of the gas content of these planets is an overestimate.

\section{Discussion}

The formation models of \cite{ormel21} stop at crossover mass, before the planet enters the runaway gas accretion phase. We show here that planets with massive envelopes (i.e., slow rainout) preserve their formation composition gradient. 
Massive (runaway) gas accretion on top of the planets we explore here would keep the deep interior undersaturated on {${\sim}\mathrm{Gyr}$} timescales. 
This result is consistent with the indications of diluted core in Jupiter and Saturn \citep{vazan16,wahl17,debras19,mankovich21}, {and implies the Z-gradient may arise} perhaps from formation{, although other mechanisms can also redistribute the metals} in the interior\footnote{In giant planets other processes such as miscibility of rock in metallic hydrogen \citep[e.g.]{wilson12b,soubiran17}, and/or convective-mixing \citep{vazan15,muller20} can change the long term interior structure.}. 

We find that the envelope mass is the key {factor that determines the timing and duration of the} silicate rainout. However, initial envelope mass is an uncertain parameter, which varies with planet formation model assumptions. The initial envelope masses found in \cite{ormel21} are {broadly} within the range of other planet formation works \citep[e.g.,][]{mordasini20,leechiang15}, but can vary with additional processes that potentially slow or halt gas accretion, like envelope recycling \citep{ormel15,kuwahara19,moldenhauer21}
that are not included in our formation model. 
Moreover, atmospheric boil-off \citep{owen16,rogers23x} can remove significant part of the envelope at very early evolution stage. 
We therefore vary the envelope masses resulting from the planet formation model, mainly to lower values than in the original model.  

{Although} the work presented here {assumes} planets are formed by pebble accretion, some of the findings may {also hold} for planets formed by planetesimal accretion. \cite{boden18} and \cite{valletta20} showed that formation of sub-Neptune planets by planetesimal accretion {also terminates} with a composition gradients, although steeper than in the pebble accretion case \citep{ormel21}. A further investigation is required to explore the silicate rainout processes in these planets.

\subsection{Model limitations}

The envelope metallicity follows the saturation L-V curve, in which silicate content increases with pressure and temperature, forming a composition gradient along the saturated envelope. For simplicity, we ignore the effect of this saturated outer composition gradient on the thermal evolution. Nevertheless, this additional composition gradient is expected to lead to convection inhibition and slow the cooling of the planet in this region \citep{guillot95,leconte17}. \cite{markham22} suggest this mechanism to lower the heat flux from deep interiors of sub-Neptunes, keeping the core in a high entropy supercritical state for billions of years, and permitting the dissolution of large quantities of hydrogen into the core. 
We anticipate this process to elongate the rainout timescales we find here, hence in this context our result provide a lower bound for the rainout timescale. 

We use the liquid-vapor curve as a sole criterion for {silicate} redistribution, as in \cite{podolak88,boden18,stevenson22}. We ignore in this work any chemical interaction between the silicates and the hydrogen. 
However, chemical interaction between species is expected to produce elements, like water and silane, that are not included in our study \citep{schaeferfeg09,misener23,horn23}. 
Dissociation of $SiO_2$ molecules in the deep interior, at temperature of about 5000\,K and above \citep{melosh07} is another mechanism that influence our results. 
Hydrogen dissolution in the rocky deep interior and out-gassing of the hydrogen as the planet cools \citep{chachan18,kite19} may affect planets with low gas content. 

{We use the \cite{freedman14} radiative opacities as a function of pressure-temperature-metallicity. The original tables are valid in the range of up to 0.03\,GPa in pressure, 4000\,K in temperature, and 50X\,solar in metallicity. For higher values we extrapolated the radiative opacity, based on the analytical expressions provided in \cite{freedman14}, as a function of pressure, temperature, and metallicity. 
The higher opacity obtained by the extrapolation leads to more convection in the outer composition gradient and upper envelope, thus, not necessarily slower cooling. Moreover, the radiative opacity effect is limited as the clean atmosphere (on top of the rainout boundary) is mainly convective, while the 
heat transport in the extrapolated high pressure layers ($>$0.1\,GPa), e.g., in the deep composition gradient is mainly conductive.
{We also note} that the \cite{freedman14} opacity is calculated for solar ratio composition, in which contribution of species to the opacity is related to their solar abundances. Atmospheric metallicity that is based on L-V curves and cloud model \citep{poser24,ormelmin19,min20} requires opacity calculation for other ratios, that may affect the rainout timescales we find here.}

The planets we study in this work are rocky (dry) planets. Our results cannot be directly applied to planets containing large fraction of volatiles (wet planets) from two main reasons. 
First, volatile evaporate at the outer atmosphere and not in the deep interior, affecting the outer envelope properties and thus the planet formation and evolution process \citep{horiikoma11,venturini16,kurosaki17,venturini20a}. 
Second, perhaps more fundamentally to our work, miscibility of rock in water at high pressure \citep{nisr20,kim21} may prohibit the {silicate} settling, leaving the rock mixed with the water on Gyr timescale \citep{vazan22,dornlich21}.
We therefore expect rainout and its consequent effects to be less remarkable in wet planets.

The thermodynamic properties of H,He are derived from the equation of state (EoS) of \cite{scvh}, for consistency with the planet formation calculation \citep{ormel21}. A newer EoS for hydrogen and helium mixture \citep{chabrier19} indicate higher compressibility than \cite{scvh} for density above 0.1 g/$cm^3$. Based on the results of \cite{chabrier19} we estimate a difference of up to 15\% in envelope density-pressures relation for the planets in this work. Consequently, thermal evolution and rainout timescale are expected to be similar for the super-Earth planets in our sample, but longer for the more massive Neptune-like planets. 
{Together with the} convective-inhibition {described above,} the {rainout timescales} presented in this work are a lower bound. 

\section{Conclusions}
By linking planet formation model to a non-adiabatic structure-evolution model, we have {calculated} the {post-formation} structure and thermal evolution of rocky planets formed by pebble accretion. {Specifically, we have followed, for the first time, the fate of the hot silicate vapor that remained in the planet's envelope after pebbles sublimated, as the planet cools}. {The model includes} heat transport by radiation, convection and conduction, and mass redistribution by rainout (condensation and settling) and by convective-mixing. 
We find that the interior structure of {silicate}-rich planets that formed with polluted envelopes can be roughly divided in three groups, based on their envelope (H,He) mass: bare rocky cores that have lost their envelopes, super-Earth planets with core-envelope structure, and Neptune-like planets with diluted cores that are slowly raining-out. 
Our {findings} bridge the gap between the indications for composition gradients in massive planets (e.g., solar system gas giants) and core-envelope structure in small planets. 
Below are our main conclusions: 

\begin{enumerate}

    \item {Formation of sub-Neptune planets by pebble accretion (and to a lesser extent, planetesimal accretion) result in a strong composition gradient within the envelope, due to the thermal ablation of the pebbles (planetesimals). This composition gradient cannot be overturned by convection, but its metals can be redistributed due to rainout.}
    
    \item Cooling of polluted envelopes leads to silicate rainout and late core growth. The interior structure of rocky sub-Neptunes is therefore determined by the time over which rainout proceeds ($t_\mathrm{rain}$), which depends mainly on the envelope {mass}, but also on planet mass and atmospheric opacity. Specific formation conditions (e.g., the pebble size or the pebble accretion rate) {are insignificant to} the rainout {process}. 
      
    \item The energy release {that accompanies} rainout -- {gravitational} (settling) {and to a lesser extent} latent heat (condensation), as well as stored formation energy -- {act to} inflate the planet radius, in comparison to a core-envelope analogue. 
    Radius inflation is inverse{ly} proportional to the rainout timescale. Lower-mass planets have shorter rainout time and larger radius inflation. We {provide an} analytical estimate for the rainout timescale and radius inflation (section~\ref{sec:anal}). {Accurately age determinations are therefore key in order to properly interpret mass-radius relationships for planets with significant gas content. }
    
    \item Young planets with envelopes lighter than 0.4\me experience their maximum radius inflation by rainout in the first hundreds of Myrs, {during which they are} prone to mass loss by photoevaporation. The mutually {enforcing processes of} radius inflation and mass loss in this period result in enhanced (sedimentation-powered) loss of {envelopes for} these planets. 

    \item Planets with massive envelopes preserve their primordial composition gradients due to the blanketing effect of the envelope. Thus, Gyr old Neptune-like planets are expected to have gradual distribution of metals in their interiors.
    
\end{enumerate}

\section*{Acknowledgments}
We thank the referee for their constructive comments. We thank Tristan Guillot, Steve Markham, Masahiro Ikoma, Jonathan Fortney, and Re'em Sari for useful discussions. 
A.V. acknowledges support by ISF grants 770/21 and 773/21. 
C.W.O. acknowledge support by the National Natural Science Foundation of China (grant no. 12250610189 and no. 12233004). 
Figures were plotted using Matplotlib \citep{matplotlib07}.


\begin{appendix}

\section{A comparison between ideal and non-ideal gas equations of state}\label{sec:app1}

A very common approximation in planet formation and evolution studies is that the gas is ideal. Its deviation from the more realistic EoS is usually not considered. One of the differences between our simulations and the analytical model in section~\ref{sec:comp} is the use of a tabular EoS in the simulations{, while in the} analytical calculation the gas is assumed ideal. To explore the difference we compare the ideal gas EoS against the tabular pressure-temperature-density relation of a H,He mixture. 
The ideal gas we used for the comparison is calculated for hydrogen and helium mass fraction of 0.7 and 0.3, respectively. The mean molecular weight of the H,He mixture is 2.35, the adiabatic index is 1.45 and the adiabatic gradient is 0.31, as in \cite{ormel21}.  

\begin{figure}
\subfigure{\includegraphics[width=9cm]{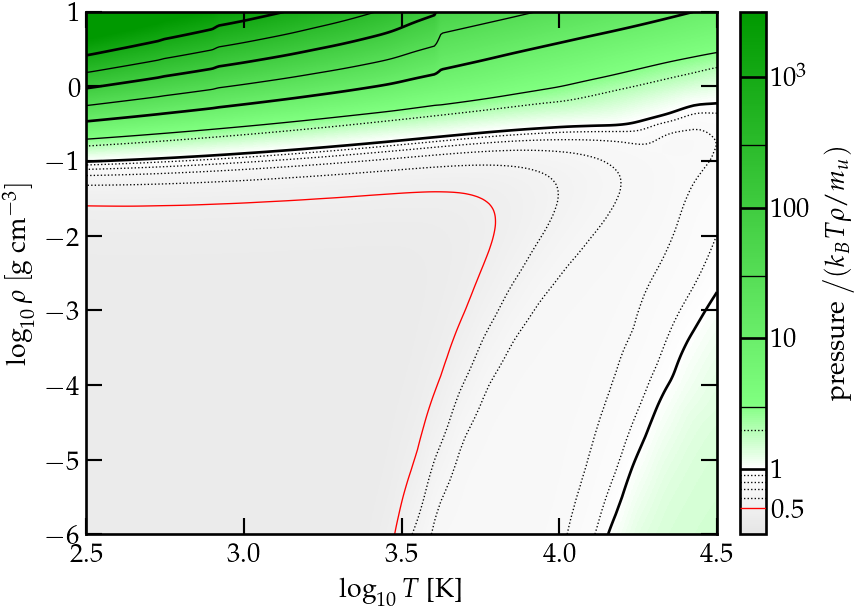}}
\subfigure{\includegraphics[width=9cm]{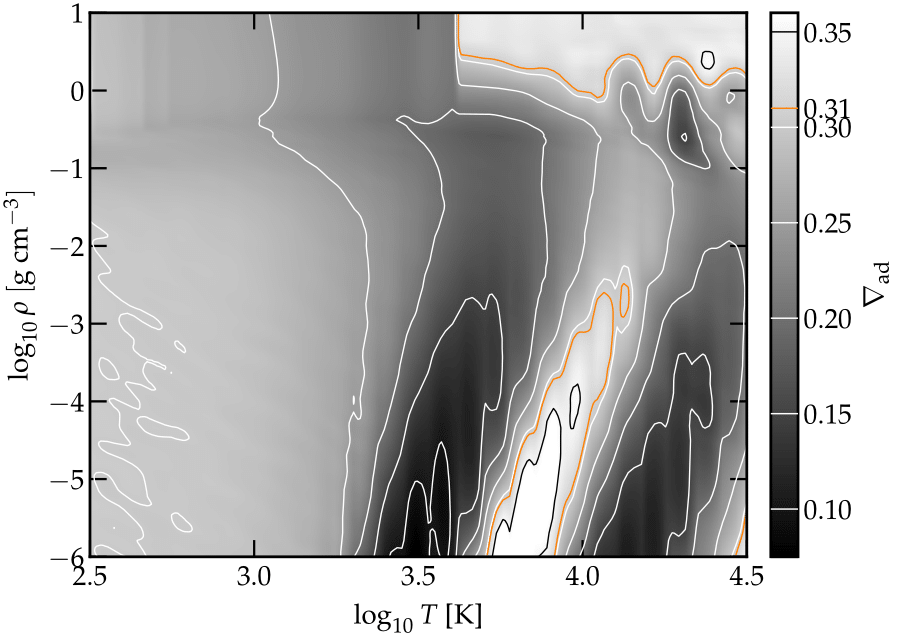}}
\caption{Hydrogen-helium mixture of 0.7 H and 0.3 He (in mass fraction), from the equations of state of \cite{scvh}. Show are pressure normalized to ideal-gas (up), and the adiabatic gradient (bottom). Red curve (up) fits the ideal gas prefactor of 1/($\mu=$2.35), the orange curve (bottom) signifies the ideal gas constant value of $\nabla_{\rm ad}=0.31$.} \label{fig:app1}
\end{figure}

In figure~\ref{fig:app1} we plot the properties of the H,He mixture as a function of density and temperature, from the tabular EoS of \cite{scvh}. The H,He mixture is calculated according to the additive volume law, as in \cite{vazan13}. Shown are the pressure normalized to the ideal gas pressure (top), and the related adiabatic temperature gradients (bottom). 
The upper panel of figure~\ref{fig:app1} indicates that gas density-pressure fairly agrees with ideal gas up to density of about 0.1 g/$cm^3$. The ideal gas pre-factor of 1/($\mu=$2.35) is shown in red curve. 
At higher density the real gas diverges from the ideal gas, and is significantly less compressible than the ideal gas. 
Pressure-temperature profiles of the planets we modelled here indicate that most of the gas is at higher than 0.1 g/$cm^3$ density for most of the evolution time. In general, sub-Neptune planets with more than a few percent of gas have large fraction of their gas at high density, in which ideal gas assumption doesn't hold. 
Consequently, modeling planets that have more than a few percent of gas by using ideal gas lead to unrealistically hotter envelopes and smaller radii.

Another phenomenon that is ignored in ideal gas calculation is the ionization and dissociation of molecules, which affects the heat capacity and thus the temperature profile and evolution. In the bottom panel of figure~\ref{fig:app1} gray-scale represents the adiabatic temperature gradient (nabla) parameter, where ideal gas value is shown in orange curve. 
The ideal gas value of 0.31 (2/7 for H and 2/5 for He) is higher than of the real gas, for gas density above 0.1 g/$cm^3$, leading to too sharp temperature-pressure slope in models that are using ideal gas.

\section{Harmonic mean of radiative and conductive opacity}\label{sec:app2}

\begin{figure}
\subfigure{\includegraphics[width=9cm]{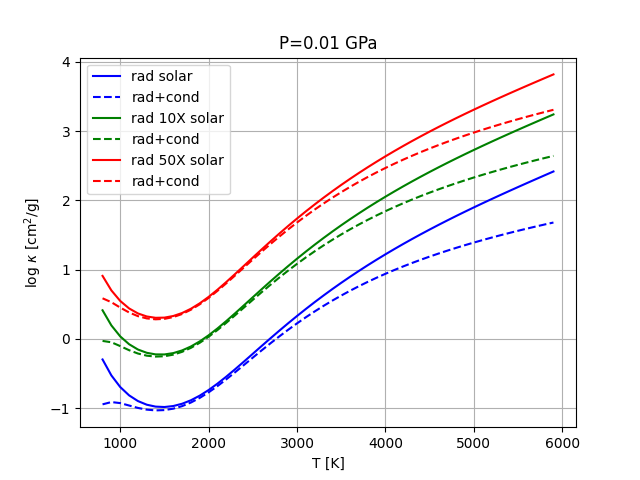}}
\subfigure{\includegraphics[width=9cm]{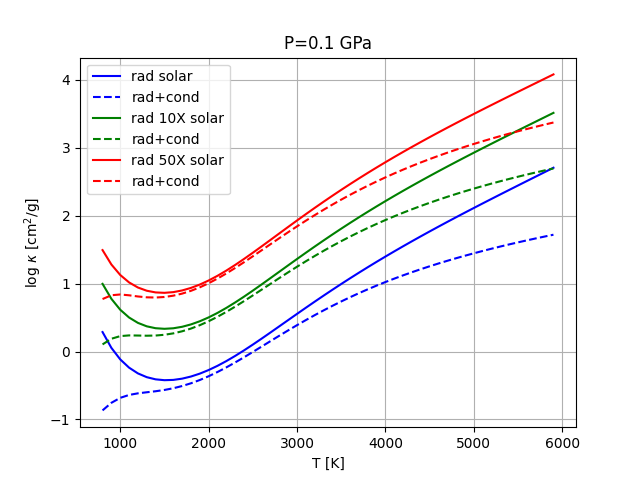}}
\subfigure{\includegraphics[width=9cm]{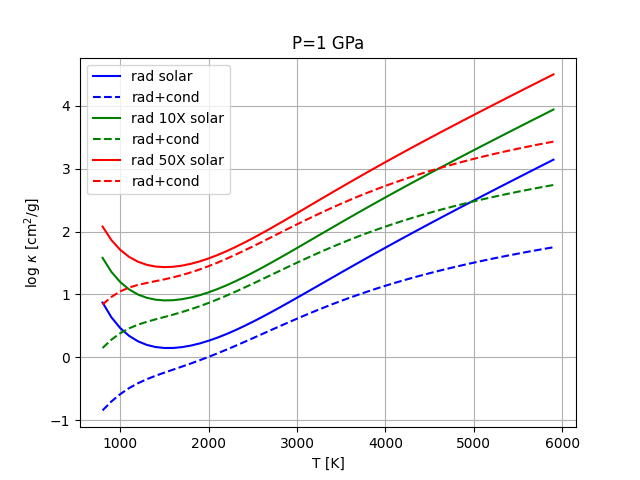}}
\caption{{Radiative opacity, based on \cite{freedman14} calculation (solid), and the harmonic mean of radiative and conduction opacity (dashed) as a function of temperature. Color represent different metallicity and each panel is for a different pressure.}} \label{fig:app2}
\end{figure}

{Conductive opacities are calculated from the thermal conductivity of silicate, as hydrogen-helium conductivity is negligible in the pressure-temperature range of interest \citep{french12}. We take the thermal conductivity of Earth, $4\,\mathrm{W\,m^{-1}\,K^{-1}}$ \citep{stevenson83} to be our reference values, and scale it with pressure and temperature according to \cite{stamenkov11}. Then, the harmonic mean of radiative and conductive opacities is calculated according to equation~\ref{eq:opac}. 
In figure~\ref{fig:app2} we present values of the radiative opacity (solid) and the harmonic mean of the radiative and conductive opacity (dashed) as a function of temperature, for three metallicity values (color) in three pressures (panels). 
Conduction scales linearly with the silicate mass fraction in a given layer. 
As can be seen, conduction dominates the heat transport as temperature and pressure increase.}

\end{appendix}


\bibliographystyle{aa}
\bibliography{allona} 


\end{document}